\documentclass[11pt]{article}

\usepackage{amssymb,amsmath,amsfonts}
\usepackage{graphicx,xcolor,enumitem}
\usepackage{epsfig}
\usepackage{amsthm}
\usepackage{bm}
\usepackage{subcaption}
\usepackage[round]{natbib}
\usepackage{csquotes}
\usepackage{float}

\renewcommand{\mkbegdispquote}[2]{\itshape}
\usepackage[a4paper, twoside, hmarginratio=1:1, vmarginratio=1:1, left=1in,top=1in]{geometry}

\RequirePackage[breaklinks=true, hidelinks]{hyperref}
\usepackage{breakcites}

\newcommand{\E}{\mathbb{E}}
\newcommand{\R}{\mathbb{R}}
\newcommand{\p}{\mathbb{P}}
\newcommand{\cL}{{\mathcal L}}

\newcommand{\cH}{{\mathcal H}}
\newcommand{\cF}{{\mathcal F}}
\newcommand{\cP}{{\mathcal P}}
\newcommand{\id}{\mathbf{1}}

\newtheorem{theorem}{Theorem}

\newtheorem{definition}[theorem]{Definition}

\newtheorem{lemma}[theorem]{Lemma}

\theoremstyle{definition}

\numberwithin{equation}{section}
\numberwithin{theorem}{section}

\begin{document}
	
\title{Cooperation between Independent Market Makers\footnote{Submitted for review on September 19, 2021.}}
\author{Bingyan Han \thanks{Previously affiliated with Division of Science and Technology, BNU-HKBU United International College, Zhuhai, Guangdong, China. The new affiliation of the author will be Department of Mathematics, University of Michigan, Ann Arbor, MI, USA. Email: byhan@umich.edu}
	}

\date{September 19, 2021}
\maketitle	
	
\begin{abstract}
	With the digitalization of the financial market, dealers are increasingly handling market-making activities by algorithms. Recent antitrust literature raises concerns on collusion caused by artificial intelligence. This paper studies the possibility of cooperation between market makers via independent Q-learning. Market making with inventory risk is formulated as a repeated general-sum game. Under a stag-hunt type payoff, we find that market makers can learn cooperative strategies without communication. In general, high spreads can have the largest probability even when the lowest spread is the unique Nash equilibrium. Moreover, introducing more agents into the game does not necessarily eliminate the presence of supra-competitive spreads.
	\\[2ex] 
	\noindent{\textbf {Keywords:} Market making, financial regulation, general-sum game, independent reinforcement learning.}
	\\[2ex]
	\noindent{\textbf {JEL Classification:} C73, D53, L40.} 
\end{abstract}

\section{Introduction}
Three decades ago, the United States stock market usually traded with a tick size of one-eighth. In 1994, \cite{christie1994nasdaq} published empirical evidence that Nasdaq market makers avoided odd-eighth quotes for 70 of 100 actively traded securities. On May 26 and 27, 1994, several newspapers reported \cite{christie1994nasdaq}'s finding that the empirical data cannot reject the hypothesis that market makers implicitly colluded to maintain spreads of at least $\$0.25$. On May 27, 1994, several dealers suddenly increased their use of odd-eighth quotes \citep{christie1994did}. These academic works led to a civil suit between tens of thousands of investors and brokerage firms. In December 1997, brokerage firms agreed to pay about $\$900$ million to end the suit, but none of them acknowledged wrongdoing. In 1997, the SEC introduced order handling rules as a structural fix for the lack of competition. Nowadays, U.S. securities markets can trade in decimals; see \cite{lindsey2016sec} for a review on related regulations.

Algorithmic trading has dramatically changed security markets. Market participants are increasingly using autonomous algorithms in their decision-making processes. However, the sophistication and powerfulness of algorithms have also led to another prominent concern on collusion. Algorithms may be sufficiently advanced to learn that it is optimal to collude \citep{ezrachi2016virtual}. Although it sounds like science fiction and almost impossible, recent experimental studies \citep{waltman2008q,klein2019autonomous,calvano2020artificial,hansen2020algorithmic} suggest that algorithms can learn collusive strategies from scratch, even without human guidance or communication with each other. \cite{calvano2020protecting} further pointed out flaws of antitrust laws on preventing algorithmic collusion.

A crucial difference between human and algorithmic collusion is, human collusion usually involves communication, which can be detected and used as evidence in the antitrust lawsuit\footnote{For the civil suit between investors and dealers, \citet[footnote 10]{christie1995policy} mentioned that there was overt harassment of dealers who broke the spreads, detailed in the {\it Los Angeles Times} on October 20, 1994 (pg. 1).}. Regulators usually identify the collusion by communication between agents, due to the concealment of collusive strategies and the difficulty in evaluating supra-competitive prices relative to marginal costs \citep{calvano2020protecting}. However, if algorithms can collude without communication, then the regulators lack the tools to stop them. An OECD Secretariat background paper\footnote{Available at {\scriptsize \url{https://www.oecd.org/competition/big-data-bringing-competition-policy-to-the-digital-era.htm}}} mentioned that ``{\it there is no legal basis to attribute liability to a computer engineer for having programmed a machine that eventually `self-learned' to coordinate prices with other machines}".

Compared with e-commerce markets, dealer markets may achieve collusion much faster due to high-frequency trading. Experiments in \cite{calvano2020artificial} converge to collusive outcomes after millions of pricing periods. This convergence speed is almost unrealistic for e-commerce since sellers typically adjust prices in several days. However, in securities trading, price adjustments are fast and millions of interactions can happen very quickly. The second difference between e-commerce and market making is that brokers commonly act as both sellers and buyers. They can implement complicated trading rules by quoting on both ask and bid sides. Inventory risk plays an essential role in market-making activities. Motivated by these crucial differences, we study the potential coordination between market makers who implement their strategies independently.

Consider a dealer market with multiple market makers and a single security. To simplify the competition between market makers, we assume all agents quote limited orders simultaneously in discrete time. The arrival of market orders is allowed to depend on the market liquidity provided by the market makers. Higher spreads reduce the trading willingness of investors. Thus the probability of execution is lower. We assume at most one order on one side can arrive at a single period. Therefore, the market order only matches the best quotes. Briefly speaking, the competition between market makers forms a repeated general-sum game. Each agent aims to maximize its payoff, including the profits from bid/ask spreads and the penalty on inventory change. The game may have multiple Nash equilibrium strategies. We also introduce the concept of cooperative strategies which maximize the joint profits of all agents. They can be thought of as the solution when all agents can coordinate perfectly with each other. Collusion in economics usually requires not only supra-competitive prices but also a punishment scheme on defection from competitors. Since we mainly consider stateless Q-learning that cannot have the memory to formulate punishment, we refer to our action profiles that lead to supra-competitive profits as cooperation instead of the legal term collusion. If agents learn independently without communication, which one will they choose among cooperative and Nash equilibrium strategies? 

This problem fits naturally in the independent reinforcement learning framework. More specifically, each agent stores a separate $Q$-value function to justify the quality of his actions. However, independent reinforcement learning is well-known for its non-stationarity. An agent's policy depends on rivals' policies, which are changing during the learning progression. This non-stationarity is a major difficulty in proving theoretical convergence guarantees. To ease the technical difficulty, we consider stateless Q-learning in theoretical proofs. Another motivation for this restriction is dark pool trading. Participants usually have limited knowledge of the current market situation, which makes the concept of states useless. 
The first contribution of this paper is a convergence theorem for a multi-agent Q-learning game, given in Theorem \ref{Thm:Contraction}. Then we consider the two-agent setting as a motivating example and obtain the following results:

\begin{enumerate}
	\item Stag hunt game with two spreads on one side (Table \ref{tab:SH}): Suppose the investors are relatively inelastic to changes in the spreads. The higher spread is sufficiently profitable such that market makers have no motivation to deviate from it. Both spreads are Nash equilibria. However, the theoretical limit suggests charging the higher spread. Experiments find algorithms can also converge to the limit that is disadvantageous to investors.
	\item Prisoner's dilemma with two spreads (Table \ref{tab:PD}): If it is optimal to deviate from the higher spread, then there exist parameter settings such that the lower spread is the theoretical long-run outcome. However, the algorithm may not converge to it if the agents favor the higher spread at first.
	\item Inventory risk has complicated effects on cooperation. It can facilitate cooperation in a stag hunt game and prevent cooperation in the prisoner's dilemma. We have considered both the $L^2$ penalty and a hard constraint formulation in experiments.
	\item Multiple bid/ask spreads: In stateless Q-learning, we find the lowest spread can be avoided considerably under a moderate temperature, even when the lowest spread is the unique Nash equilibrium. However, the temperature constant is crucial and a smaller one leads to outcomes favoring the lowest spread. 
	\item As an extension, we consider the Q-learning with memory experimentally. Far-sighted market makers with a large discount factor charge higher spreads more frequently, even under multiple bid/ask spreads and the small temperature setting. However, corresponding theoretical guarantees are left open.
\end{enumerate}

\begin{table}[!h]
	\centering
	\begin{tabular}{c | c | c}
		\hline
		Spreads & 0.1 & 0.8 \\ 
		\hline
		0.1 & (0.05, 0.05)  & (0.1, 0) \\
		\hline
		0.8 & (0, 0.1) & (0.4, 0.4) \\
		\hline
	\end{tabular}

	\caption{Payoff matrix under a stag hunt game. Two market makers (agents) ignore inventory risk and quote on one side only. The payoff tuple is for (agent 1, agent 2). Agent 1 selects spreads in the row and agent 2 selects in the column. Market orders always arrive with probability one.} \label{tab:SH}
\end{table}

\begin{table}[!h]
	\centering
	\begin{tabular}{c | c | c}
		\hline
		Spreads & 0.41 & 0.8 \\ 
		\hline
		0.41 & (0.205, 0.205)  & (0.41, 0) \\
		\hline
		0.8 & (0, 0.41) & (0.4, 0.4) \\
		\hline
	\end{tabular}
	\caption{ Payoff matrix with a prisoner's dilemma. Market orders arrive with probability one.}\label{tab:PD}
\end{table}

The first idea in mind to prevent cooperation is to introduce more market makers into the competition. However, we theoretically prove that the limit is not necessarily the lowest spread, even with an infinite number of agents. In other words, including more agents is not an effective way to destabilize cooperation between agents.

The rest of the paper is organized as follows. Section \ref{Sec:Market} presents the dealer market. Section \ref{Sec:InQ} describes the Q-learning framework for market making. Motivating examples with two agents are given in Section \ref{Sec:TwoAgents}. The code is publicly available at \url{https://github.com/hanbingyan/CoinMM}. Section \ref{Sec:InfAgents} shows the theoretical limit when the number of agents goes to infinity. Section \ref{Sec:Dis} concludes the paper with a discussion on future research directions. Technical proofs are given in the Appendix.

\section{Dealer market}\label{Sec:Market}
Consider a dealer market with $N$ market makers and a single traded stock. Market makers provide the prices at which they are willing to buy (bid) and sell (ask). Commonly, there is also an exchange limit order book (LOB) market for the same security, which provides reference pricing information. The mid-price $S_t$ at time $t$ is the mean value of the best bid and ask in the LOB. Market makers also quote their bid and ask spreads relative to $S_t$. In this paper, we assume bid/ask spreads are quoted discretely from a set $\{K(1), ..., K(M)\}$ with $M$ values in ascending order. At time $t$, for market maker $i$, denote his ask spread index as $a_i(t)$ and bid spread index as $b_i(t)$, where $a_i(t)$ and $b_i(t)$ belong to $\{1, 2, ..., M\}$. Let $K(a_i(t))$ and $K(b_i(t))$ be the corresponding ask and bid spreads. Then the market maker $i$ is willing to buy at $S_t - K(b_i(t))$ and sell at $S_t + K(a_i(t))$. If a trade of size $n$ is executed between the market maker $i$ and an investor at time $t$, then the market maker earns a profit of $n K(a_i(t))$ when it is a sell and $n K(b_i(t))$ when it is a buy\footnote{Buys and sells are from the market maker's perspective.}. In practice, bid/ask spreads are functions of trade sizes. However, for simplicity, we assume that the time step can be split small enough such that the trade size for one side is at most one, inspired by \cite{baldacci2021optimal}. But different with \cite{baldacci2021optimal}, we suppose that the time step is also discrete, as in \cite{nips19MM}.

For later use, combine the bid/ask spreads for the market maker $i$ as $c_i(t) := (a_i(t), b_i(t))$. Following the terminology of Q-learning, we also refer to $c_i(t)$ as an action. The action space size is therefore $M^2$. Besides, aggregate actions of all agents in one period as the following vectors
\begin{align*}
A(t) := (a_1(t), ..., a_N(t)), \quad B(t) := (b_1(t), ..., b_N(t)), \quad C(t) := (c_1(t), ... , c_N(t)).
\end{align*}  
Let $\underline{A}$ be the best ask spread in $A$ and $|\underline{A}|$ be the number of agents quoting the best ask. The notations for the bid side are defined similarly.

There are mainly two types of orders in the market. Limit orders set the maximum or minimum price at which a participant is willing to buy or sell and the order is executed only when the price is matched. Market orders are executed immediately with the best price available. We assume market makers always use limit orders and investors use market orders. There exist numerous models for market order arrival intensities such as \cite{avellaneda2008high,baldacci2021optimal}. The arrival probabilities of market orders depend on the market liquidity provided by market makers. In period $t$, denote the arrival probability of a market order on the ask side as $\cP(A(t)) := e^{- f(A(t))}$. Similarly, bid side arrival probability is $\cP(B(t)) := e^{- f(B(t))}$. We use the same deterministic function $f(\cdot)$ here only for simplicity. Our framework allows a general $f$ to capture stylized facts in the market. For example, $\cP(\cdot)$ can be a decreasing function of the ratio between the spread and the volatility, as specified in \eqref{Eq:ArrivalP}.

Since we assume at most one market order for each side can arrive during a single period, only the best quotes can get the market order. If there are multiple winners, we assume they split the order equally. After one period, suppose all remaining limit orders are revised with new prices unless it is filled. This assumption is not restrictive, since a considerable amount of limit orders are finally revised or canceled in practice. 

Besides the profits from bid/ask spreads, market makers bear the risk from inventory in the stock, due to mid-price fluctuations. Denote the accumulated inventory of agent $i$ after trades at time $t$ as $y_i(t)$. The calculation of the inventory is straightforward. If the market maker $i$ receives $a$ ask orders and $b$ bid orders in period $t$, then the accumulated inventory is $y_i(t) = y_i(t-1) + b - a$. Suppose $y_i(-1) = 0$.

The total reward $r_i(t)$ in period $t$ has three parts
\begin{equation*}
r_i(t) : = r_{i, a}(t) + r_{i, b}(t) - \xi (y_i(t) - y_i(t-1))^2.
\end{equation*}
The first part $r_{i, a}(t)$ is from the ask side, with an expected value of
\begin{equation}\label{Eq:ExpectedRa}
\E[r_{i, a}(t)|A(t)] = K(a_i(t)) \id_{\{ a_i(t) = \underline{A(t)} \}} \frac{1}{|\underline{A(t)}|} \cP(A(t)). 
\end{equation} 
Similarly, for the bid side,  
\begin{equation*}
\E[r_{i, b}(t)| B(t)] = K(b_i(t)) \id_{\{ b_i(t) = \underline{B(t)} \}} \frac{1}{|\underline{B(t)}|} \cP(B(t)). 
\end{equation*} 
We deduct the third quadratic cost from the profits to model the inventory risk. Constant $\xi$ represents the aversion to inventory risk and is assumed to be the same for all market makers for simplicity.

In summary, if we know all agents' actions $C(t)$, then the conditional expectation of $r_i(t)$ is
\begin{equation*}
\E [r_i (t)| C(t) ] =  \E[r_{i, a}(t)|A(t)] + \E[r_{i, b}(t)| B(t)] - \xi  \E[(y_i(t) - y_i(t-1))^2| C(t) ].
\end{equation*}
Compared with the reward mechanism in \cite{calvano2020artificial}, undercutting spreads incurs more severe harms in our formulation since higher spreads will not receive the order. From this perspective, it is harder to achieve cooperation under our design. 

Each agent aims to maximize its cumulative profit $\E[ \sum^\infty_{t=0} \gamma^t r_i(t)]$, with no knowledge on other agents' selections and rewards. $\gamma$ is a constant discount factor. Since market makers quote new ask and bid spreads every period, market making is then formulated as a repeated game.  

To proceed, we need some game theory concepts to describe the properties of policies \citep{Nowe12}. It is well-known that the multi-agent game is non-stationary and the convergence is an open problem in general. Since we aim to investigate the convergence and limiting behaviors, a restrictive assumption needed is that the game is stateless, as in \cite{waltman2008q,Wunder10}. However, this assumption is not fully unrealistic for market making. Participants in dark pool trading usually have limited knowledge of the states of the market environment.

Under the stateless setting, the expected payoff $\E [r_i | C]$ to agent $i$ is independent of periods $t$. We can consider static payoff matrices for market makers. The game in one period becomes a matrix game. A random strategy, also called a {\it mixed} strategy, is a probability distribution on $M^2$ actions. Deterministic strategies are special cases that assign probability 1 to a single action and are also called {\it pured}. See \cite{Nowe12} for a survey on related concepts. In a Nash equilibrium, each agent acts the best response to other agents' choices.

\begin{definition}
	$C^* = (c^*_1, ..., c^*_N)$ is a pure Nash equilibrium if
	\begin{equation}
	\E [r_i | C^*] \geq \E [r_i | C^*_{-i}, c_i]
	\end{equation}
	for all agents $i$ and action $c_i$, where $C^*_{-i}$ is $C^*$ except $c^*_i$.
\end{definition}
In other words, if all other agents $j$, with $j \neq i$, choose $c^*_j$, then it is optimal for agent $i$ to choose $c^*_i$. An arbitrary matrix game may not have a pure Nash equilibrium, even for a two-agent case. However, a two-agent general-sum game always has a mixed Nash equilibrium \citep{Nowe12}. Besides, a matrix game may also have multiple pure Nash equilibria. 

\begin{definition}
	$\hat{C}^* = (\hat{c}^*_1, ..., \hat{c}^*_N)$  is a cooperative strategy if it maximizes $\E [ \sum^N_{i=1} r_i | C]$, the joint profits of all agents.
\end{definition}         
It is optimal for all agents to select a cooperative strategy $\hat{C}^*$. However, if agents have motivation to cheat and deviate from $\hat{C}^*$, they may finally result in a Nash equilibrium with much lower profits. If all agents use algorithms to learn independently, do they always converge to a Nash equilibrium? If there are multiple Nash equilibria, which one do they choose? To answer these questions, we consider a classical Q-learning framework.

\section{Independent Q-learning}\label{Sec:InQ}
Since we consider finite action spaces and forbid communication between market makers, then independent Q-learning \citep{claus1998dynamics,Wunder10} is suitable for the repeated game. Each agent maintains a separate $Q$-value function, denoted as $q_i(c_i)$ for a generic action $c_i$. To approximate the unknown $Q$-value function, starting from an arbitrary initial $q_{i, 0}$, Q-learning updates $Q$-values by
\begin{equation}\label{Eq:update}
q_{i, t+1} (c_i(t)) = (1 - \alpha_t) q_{i, t} (c_i(t)) + \alpha_t [ r_i(t) + \gamma \max_{c'_i} q_{i,t}(c'_i)].
\end{equation}
$c_i(t)$ is the action of agent $i$ in period $t$. $0 \leq \alpha_t \leq 1$ is the learning rate. $r_i(t)$ is the single period reward for agent $i$. It also depends on other agents' actions, which makes the problem non-stationary.  To establish theoretical convergence results for multi-agents, we restrict to the stateless setting. However, it is direct to consider the algorithms with states experimentally in Section \ref{sec:mem}, while the convergence proof is left open. We find the stateless formulation has already demonstrated insights in the cooperation problem.

Suppose agent $i$ follows Boltzmann action selection mechanism and chooses action $c_i$ with a probability given by
\begin{equation}\label{Eq:Boltz}
\frac{e^{q_i(c_i)/\lambda}}{\sum_{c'}  e^{q_i(c')/\lambda} } =: \cL(q_i, c_i).
\end{equation}
Constant $\lambda > 0$ is usually called the temperature. We assume all agents use the same $\lambda$ for simplicity. Note that the reward $r_i(t)$ is bounded in this paper and $Q$-values  updated by \eqref{Eq:update} are thus bounded. If $\lambda$ approaches infinity, then market makers select all actions with equal probability. If $\lambda \to 0$, then agents select the actions with the highest $Q$-values. For any fixed $\lambda > 0$, random strategy \eqref{Eq:Boltz} satisfies 
\begin{equation*}
\p[ c_i(t) = w ] > 0 
\end{equation*}
for all actions $w$. Thus, each action can be visited infinitely.

We aim to study the long-run behavior of all market makers, with the updating rule \eqref{Eq:update} and Boltzmann selection mechanism \eqref{Eq:Boltz}. Concatenate $Q$-values of all agents as a single vector $Q_t := (q_{1,t}, ..., q_{N,t})$. We introduce an operator $\cH[\cdot]$ such that 
\begin{align*}
\cH [q_i(w)] =& \sum_{C |_{c_i = w}} \left\{ \Big[\prod_{ j \neq i} \cL (q_j, c_j) \Big] \E[r_i | C] \right\} + \gamma \max_{w'} q_i(w') \\
=: & R_i(w; Q_{-i}) + \gamma \max_{w'} q_i(w').
\end{align*}
$R_i(w; Q_{-i})$ is the expected reward for agent $i$ when he selects action $w$. The summation is over all possible actions of other agents $j$, while action $c_i$ for agent $i$ is fixed as $w$. $R_i(w; Q_{-i})$ depends on all other agents' $Q$-values through probabilities $\cL(q_j, c_j)$. However, it does not depend on $q_i$ since action $w$ is chosen. $Q_{-i}$ denotes all $Q$-values except agent $i$'s. When we apply $\cH$ dynamically, it maps an $NM^2$-dimensional vector $Q_t$ to $NM^2$-dimensional $Q_{t+1}$.  

Since we assume all agents use the same temperature $\lambda$ and the game setting is also symmetric, we can expect $q_{i,t}(w)$ converges to the same $q^*(w)$ for all agents $i$, defined as 
\begin{align}
q^*(w) & = \sum_{C |_{c_1 = w}} \left\{ \Big[\prod_{j \neq 1} \cL (q^*, c_j) \Big] \E[r_1 | C] \right\} + \gamma \max_{w'} q^*(w') \nonumber \\
& = R_1(w; q^*) + \gamma \max_{w'} q^*(w'). \label{Eq:q*}
\end{align} 
Note that $i=1$ is used as an illustration. \eqref{Eq:q*} is the same for any other agents $i$. 

To prove that $q_{i,t}(w)$ converges to $q^*(w)$, we show that $\cH$ is a contraction operator under certain conditions and $q^*$ is a fixed point of $\cH$. Lemma \ref{Lem:Contract} studies the property of $\cH$ with respect to the supremum norm.

\begin{lemma}\label{Lem:Contract}
	For any $Q$-values vectors $Q$ and $Q'$,
	\begin{equation}\label{Eq:Contract}
	|| \cH[Q] - \cH[Q'] ||_\infty \leq \big( (N-1)M^2 \max_{i, w} \frac{|R_i(w; \eta_{-i})|}{\lambda} + \gamma \big) || Q - Q'||_\infty,
	\end{equation}
	where $\eta$ is a point on the line segment with $Q$ and $Q'$ as endpoints.
\end{lemma}

Note that $R_i(w; Q_{-i})$ is bounded for any $i$, $w$, and $Q$. It also relies on the number of agents $N$ implicitly since $N$ will impact the profits of a single agent. The coefficient on the right-hand side of \eqref{Eq:Contract} depends on the number of agents and the number of feasible actions. It agrees with the intuition that the operator violates contraction property more commonly when there are more agents and more choices on actions. It is useful to view $\frac{|R_i(w; \eta_{-i})|}{\lambda}$ as a whole and we will consider its limiting behavior later. Crucially, $\lambda$ cannot be too small. A counterexample is given in Section \ref{Sec:TwoAgents} to show that the algorithm does not converge to the fixed point if $\lambda$ is too small. Constant $N-1$ outside reflects that during the learning, we usually do not have same $Q$-values $q_i(w)$ for all agents $i$. Therefore, we have to include them as $N$ different vectors. Constant $\gamma$ also appears in the single-agent setting.

Theorem \ref{Thm:Contraction} proves the convergence of $q_{i,t}$ to $q^*$ with the help of Lemma \ref{Lem:Contract}. The proof idea follows from the classic result of \cite{jaakkola1994convergence}, based on stochastic approximation.
\begin{theorem}\label{Thm:Contraction}
	Suppose 
	\begin{enumerate}
		\item $0 \leq \alpha_t \leq 1$, $\sum_{t} \alpha_t = \infty$, and $\sum_t
		\alpha^2_t < \infty$; 
		\item there exists a constant $d<1$ such that $(N-1)M^2 \max_{i, w} \frac{|R_i(w; \eta_{t, -i})|}{\lambda} + \gamma \leq d$ for all $\eta_t$ on the line segments between $Q_t = (q_{1,t}, ... , q_{N, t})$ and $Q^* := (q^*, ..., q^*)$, $t = 0, 1, ...$.
	\end{enumerate}
	Then $q_{i,t}(w)$, updated by \eqref{Eq:update} with Boltzmann selection \eqref{Eq:Boltz}, converges to $q^*(w)$ in \eqref{Eq:q*} with probability 1.
\end{theorem}

\section{Motivating example: Two agents}\label{Sec:TwoAgents}
For simplicity, consider a discount factor $\gamma = 0$ in this section as an illustration. Then 
\begin{align}\label{Eq:Q_gam0}
q^*(w) & = \sum_{C |_{c_1 = w}} \left\{ \Big[\prod_{j \neq 1} \cL (q^*, c_j) \Big] \E[r_1 | C] \right\}  = R_1(w; q^*).
\end{align} 
The long-run probability satisfies
\begin{align}\label{Eq:Prob-TwoAgents}
\cL(q^*, w) = \frac{ \exp(\frac{R_1(w; q^*)}{\lambda}) }{\sum_{w'}  \exp(\frac{R_1(w'; q^*)}{\lambda}) }.
\end{align}

Motivated by \cite{avellaneda2008high,baldacci2021optimal}, consider the market order arrival probability $\cP(A)$ as a decreasing function of the ratio between the spread and the volatility. A specification of the arrival probability is 
\begin{equation}\label{Eq:ArrivalP}
\cP(A) = e^{-f(A)} = \exp \left( - \frac{\sum^M_{i=1} \bar{\omega}_i v_i}{\sigma N} \right).
\end{equation}
For a given ask spreads vector $A$, $v_i$ is the number of agents choosing spread $i$.  $v_i/N$ is then the fraction of market makers choosing spread $i$. $\bar{\omega}_i$ is a non-negative weighting factor and non-decreasing in $i$. Thus higher spreads reduce the arrival probability more severely. $\sigma$ is the volatility of the mid-price $S_t$.

To understand the dynamics of this market-making game, we start from the two-agent general-sum game, which has a long history. For example, with a dynamical system approach, \cite{Wunder10} considered a two-agent two-action game with $\varepsilon$-greedy exploration. They also identified several classes of problems that result in payoffs higher than Nash equilibrium. 

\subsection{Two spreads without inventory risk}

First, we consider the case without inventory risk, i.e., $\xi = 0$. Intuitively, decisions on the ask and bid sides should be separable. We will verify this fact later in Lemma \ref{Lem:Sep}. Take the ask side problem as an example. Theorem \ref{Thm:Contraction} is directly applicable with $M$ ask spreads, replacing $M^2$ combined actions in total. 

Since there are two agents only, denote $ z_{kl} = \E[r_{1,a} | a_1 = k,  a_2 = l]$ as the expected payoff for agent 1, when agent 1 selects ask spread $k$ and agent 2 selects ask spread $l$. Denote $p = (p_1, ..., p_M)$ as the long-run probability for $M$ ask spreads. Note that with $N=2$, $R_1(i; q^*_a) = \sum^M_{l=1} p_l z_{il}$, where $q^*_a$ is the $Q$-values in \eqref{Eq:Q_gam0} for ask spreads only. Then \eqref{Eq:Prob-TwoAgents} satisfies
\begin{align*}
	\lambda \sum^M_{k = 1} p_k \ln \frac{p_i}{p_k} =& R_1(i; q^*_a) - \sum^M_{k=1} p_k R_1(k; q^*_a) =  \sum^M_{l=1} p_l z_{il}  - \sum^M_{k=1} p_k \left(\sum^M_{l=1} p_l z_{kl} \right).
\end{align*}

In the case with two spreads ($M=2$), we have
\begin{align*}
& \lambda p_1 \ln \frac{p_1}{p_1} + \lambda (1-p_1) \ln \frac{p_1}{1-p_1} \\
& = z_{11}p_1 + z_{12} (1 - p_1) - p_1[ z_{11}p_1 + z_{12} (1 - p_1)] - (1-p_1)[ z_{21}p_1 + z_{22} (1 - p_1)] \\
& = (1- p_1)[ z_{11}p_1 + z_{12} (1 - p_1) - z_{22}(1-p_1)], 
\end{align*}
where we used the fact that $z_{21} = 0$. Then
\begin{align}\label{Eq:linear}
\ln \frac{p_1}{1-p_1} = \frac{z_{12} - z_{22}}{\lambda} + \frac{z_{11} - z_{12} + z_{22}}{\lambda} p_1.
\end{align}

\begin{figure}
	\centering
	\includegraphics[width=0.45\linewidth]{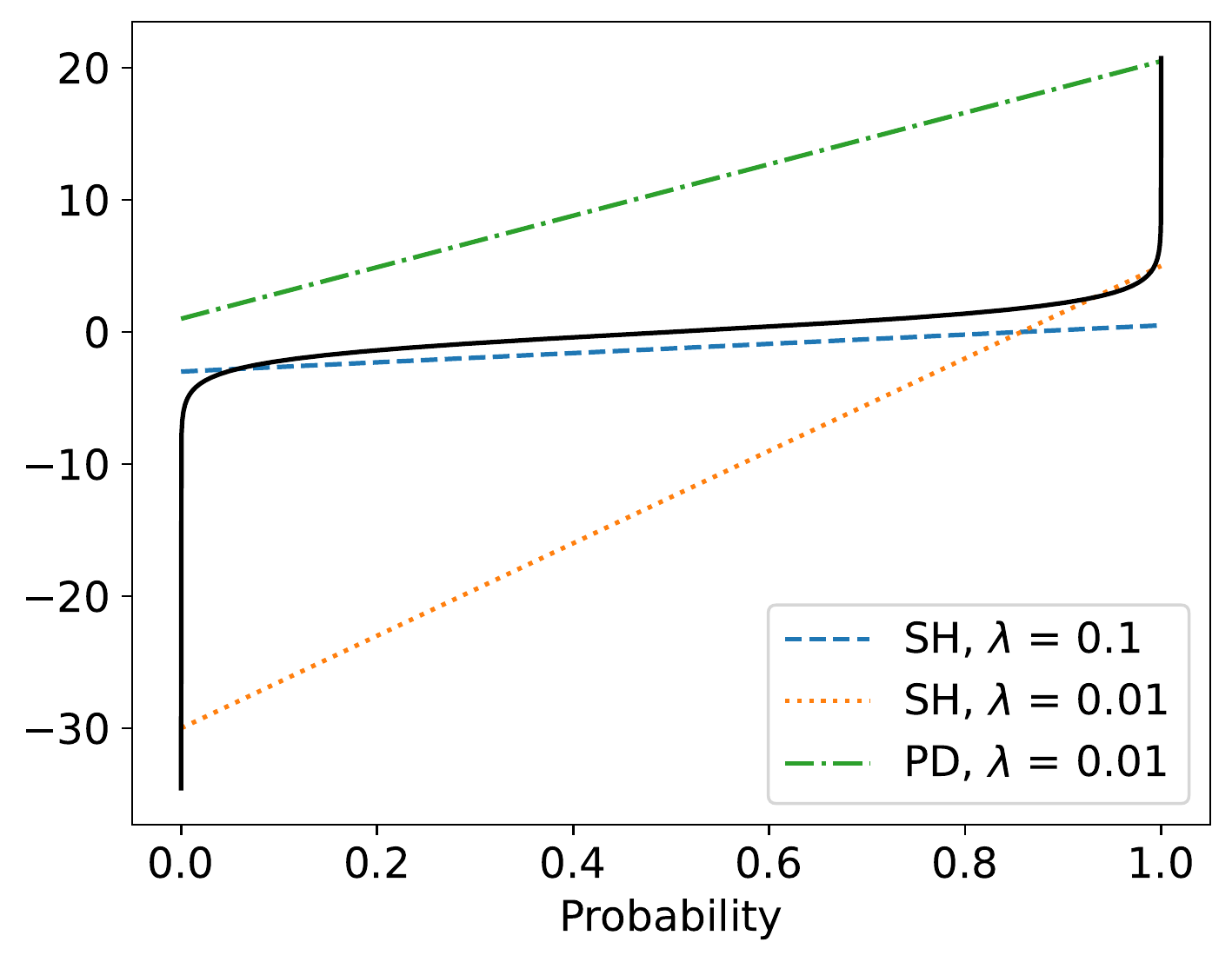}
	\caption{The stag hunt and prisoner's dilemma games have distinct theoretical probabilities of the lower spread, illustrated as crossing points. The black solid curve represents $\ln[p_1/(1-p_1)]$. Straight lines are for Table \ref{tab:SH} as an SH game and Table \ref{tab:PD} as a PD game. Crossing points between the curve and the line indicate theoretical $p_1$, the probability of choosing the lower spread, under different settings.}
	\label{fig:TwoAgents}
\end{figure}

Denote the linear function of $p_1$ on the right-hand side as $F(p_1)$. It has clear economic meanings. Recall the expected reward \eqref{Eq:ExpectedRa} and the arrival probability \eqref{Eq:ArrivalP}, then  
\begin{align*}
z_{11} + z_{22} - z_{12} = & \frac{K(1)}{2} e^{-\frac{\bar{\omega}_1}{\sigma}} + \frac{K(2)}{2} e^{-\frac{\bar{\omega}_2}{\sigma}} - K(1) e^{ - \frac{\bar{\omega}_1 + \bar{\omega}_2}{2 \sigma}} \\
 > & \frac{K(1)}{2} e^{-\frac{\bar{\omega}_1}{\sigma}} + \frac{K(1)}{2} e^{-\frac{\bar{\omega}_2}{\sigma}} - K(1) e^{ - \frac{\bar{\omega}_1 + \bar{\omega}_2}{2 \sigma}} \\
 = & \frac{K(1)}{2} \big( e^{-\frac{\bar{\omega}_1}{2\sigma}} - e^{-\frac{\bar{\omega}_2}{2\sigma}} \big)^2 > 0.
\end{align*}
Therefore, the slope of $F(p_1)$ is always positive. When $p_1$ is close to $1$, $F(p_1)$ approaches $z_{11}/\lambda$, which is also positive. The intercept depends on $z_{12} - z_{22}$. If it is more profitable for agent 1 to undercut his opponent, i.e., $z_{12} > z_{22}$, then the intercept is positive. Note the shape of $\ln \frac{p_1}{1 - p_1}$ in Figure \ref{fig:TwoAgents}. Then $p_1$, the probability for the lower spread, is always higher than $50\%$. If $z_{12} < z_{22}$, then we can find a temperature $\lambda$ such that $p_1$ is lower than $50\%$ or even close to zero. It implies that the agent will select the higher spread, if his opponent has selected it.

Table \ref{tab:SH} provides an example with $z_{12} < z_{22}$, which belongs to the so-called stag hunt (SH) game. Suppose the agents focus on two spreads $0.1$ and $0.8$. Investors have strong trading demand and are insensitive to the changes in spreads. $\bar{\omega}_1 = \bar{\omega}_2 = 0$. Then market orders always arrive with probability one. When both agents select the spread $0.8$, they have to split the single order and receive a profit of $0.4$. If one agent selects $0.1$ to undercut the competitor, he will receive the full order with a profit of $0.1$, which is not attractive. However, if they both charge $0.1$ and receive a profit of $0.05$, they will also not deviate from it since a higher spread will not receive any order. Therefore, both $(0.1, 0.1)$ and $(0.8, 0.8)$ are Nash equilibria, while the later is disadvantageous to the investors. Apparently, $(0.8, 0.8)$ is also the cooperative strategy. 

When the transactions demand of the investors is relatively inelastic to changes in the spreads, then the payoff matrix in Table \ref{tab:SH} becomes possible. Higher spreads do have higher profits and there is also no motivation for them to deviate. \cite{christie1995policy} anticipated that this insensitivity of investors to raises in trading costs is also one reason that market makers avoid odd-eighths.

Figure \ref{fig:TwoAgents} shows theoretical values of $p_1$. The solid black curve corresponds to the left-hand side of \eqref{Eq:linear}. Straight lines are the right-hand side $F(p_1)$. In our experiments under Table \ref{tab:SH}, we first set temperature $\lambda = 0.1$. The theoretical $p_1$, where the blue dashed line meets the black curve in Figure \ref{fig:TwoAgents}, is close to zero. Agents prefer the cooperative strategy in theory. Experimentally, let the learning rate $\alpha_t = 10^4/(10^4 + t)$ and the initial $Q$-values be zero. Agents charge each spread with equal probability at the beginning. Table \ref{tab:converge} shows that our training instances converge to the theoretical choice $(0.8, 0.8)$. Long-run probabilities in experiments are close to theoretical values. It demonstrates that market makers can learn cooperative strategies even with independent algorithms. 

Note that the choice of $\lambda=0.1$ with Table \ref{tab:SH} does not exactly obey the contraction property \eqref{Eq:Contract}. \eqref{Eq:Contract} is indeed a sufficient condition instead of necessary. However, if $\lambda$ deviates from \eqref{Eq:Contract} too much, two issues will arise. First, the theoretical solution $p_1$ can be non-unique. Take $\lambda = 0.01$ as an example. The orange dotted line in Figure \ref{fig:TwoAgents} meets the curve in multiple points. The probability of the lower spread can be close to either zero or one. In practice, the long-run outcomes of the algorithm rely on the initial $Q$-values and the specification of learning rates. We provide two examples in Table \ref{tab:converge}. If the agents favor the lower spread $0.1$ at the beginning by assigning a higher $Q$-value, and the learning rate converges to zero too fast, then the long-run outcome will be the spread $0.1$. In contrast, with zero initial $Q$-values and slower converging learning rates, the long-run outcome is the higher spread. Therefore, both limits are achievable in practice. Secondly, even when the limit is unique, the algorithm may not converge to it. This claim is verified soon below.

Table \ref{tab:PD} gives an example when it is optimal to deviate from the higher spread, which belongs to a Prisoner's Dilemma (PD). Table \ref{tab:PD} assumes market orders arrive with probability one. There are two spreads, $0.41$ and $0.8$. If one agent selects $0.41$ to undercut the competitor, he will receive the full order and earn a profit of $0.41$. Agents then have the motivation to deviate from the higher spread. When we select $\lambda = 0.01$, $p_1$, the probability of the spread $0.41$, is close to one, shown in Figure \ref{fig:TwoAgents} as the point where the green dash-dotted line meets the black curve. $\lambda = 0.01$ is relatively small in this setting. In practice, since the contraction property is likely to be violated, independent Q-learning does not necessarily converge to the theoretical limit for any initial $Q_0$. Indeed, suppose we consider $q_i(1) = 0$ and $q_i(2) = 0.02$. Initially, the agents prefer the higher spread. Let the learning rate $\alpha_t = 10^4/(10^4 + t)$. Experimentally, we observe that 8 out of 10 instances converge to the higher spread, which is against the theoretical result due to the violation of the contraction property. Thus, it presents a second possible source to converge to a cooperative strategy. Different from Table \ref{tab:SH}, agents demonstrate the willingness to cooperate and charge higher spreads in this situation. Besides, if we use zero initial $Q$-values, the long-run outcome agrees with the theoretical limit for 8 out of 10 instances. 

\begin{table}[H]
	\centering
	\small
	\begin{tabular}{c  c  c  c  c}
		\hline
		Setting & $\lambda$ & Initial $Q$ & Learning rate $\alpha_t$ & Long-run probabilities\\ 
		\hline
		Table \ref{tab:SH} & 0.1  & (0, 0) & $ \frac{10^4}{10^4 + t}$ & (0.0563, 0.9437) \\
		\hline
		Table \ref{tab:SH} & 0.01 & (0, 0) & $ \frac{10^4}{10^4 + t}$ & (0.0, 1.0)\\
		\hline
		Table \ref{tab:SH} & 0.01 & (0.02, 0) & $ \frac{100}{100 + t}$ & (0.991, 0.009) \\
		\hline
		Table \ref{tab:PD} & 0.01 & (0, 0) & $ \frac{10^4}{10^4 + t}$ & (0.9998, 0.0002)\\
		\hline
		Table \ref{tab:PD} & 0.01 & (0, 0.02) & $ \frac{10^4}{10^4 + t}$ & (0.0, 1.0) \\
		\hline
	\end{tabular}
	\caption{Experiments with payoffs in Tables \ref{tab:SH} and \ref{tab:PD}. Long-run probabilities are implied by mean $Q$-values of ten instances. Algorithms terminate when greedy policies remain unchanged for one million periods.}	\label{tab:converge}
\end{table}

\subsection{Two spreads with inventory risk}
Next, we consider the effect of inventory risk by setting constant $\xi > 0$. First, inventory risk reduces the probability of skewed actions on the two sides since they incur a larger inventory change. Agents prefer balanced actions with equal ask and bid spreads under the symmetric environment. Second, higher spreads become more attractive since they will receive the orders less likely, which leads to zero inventory change. Overall, inventory risk has complicated interactions with the level of cooperation. Table \ref{tab:inven} calculates the theoretical probabilities with different inventory risk aversion, under the settings in Tables \ref{tab:SH} and \ref{tab:PD}. As expected, skewed actions $2$ and $3$ have lower probabilities in both cases when $\xi$ is larger. However, inventory risk facilitates the cooperation under the SH game in Table \ref{tab:SH}, while it increases the probabilities of lower spreads in the PD game depicted by Table \ref{tab:PD}. Inventory risk seems to make the market makers focus on the actions with the highest probability since they incur no inventory penalty. 

\begin{table}[H]
	\centering
	\small
	\begin{tabular}{c  c  c  c c}
		\hline
		Table \ref{tab:SH} Actions(Ask, Bid) & 1: (0.1, 0.1) & 2: (0.1, 0.8) & 3: (0.8, 0.1) & 4: (0.8, 0.8) \\ 
		\hline
		$\xi$ = 0 & 0.00329 & 0.05408 & 0.05408 & 0.88856 \\
		\hline
		$\xi$ = 0.1  & 0.00309 & 0.04142 & 0.04142 & 0.91407 \\
		\hline
		$\xi$ = 0.2  & 0.00295 & 0.03183 & 0.03183 & 0.93338 \\
		\hline
		\hline
		Table \ref{tab:PD} Actions(Ask, Bid) & 1: (0.41, 0.41) & 2: (0.41, 0.8) & 3: (0.8, 0.41) & 4: (0.8, 0.8) \\ 
		\hline
		$\xi$= 0 & 0.72903 & 0.12480 & 0.12480 & 0.02137 \\
		\hline
		$\xi$ = 0.1 & 0.78150 & 0.09893 & 0.09893 & 0.02065 \\
		\hline
		$\xi$ = 0.2 & 0.82420 & 0.07789 & 0.07789 & 0.02001 \\
		\hline
	\end{tabular}
	\caption{Theoretical probabilities for actions under inventory risk. Temperature $\lambda = 0.1$.}	\label{tab:inven}
\end{table}

\begin{figure}[H]
	\centering
	\begin{minipage}{0.49\textwidth}
		\centering
		\includegraphics[width=0.95\textwidth]{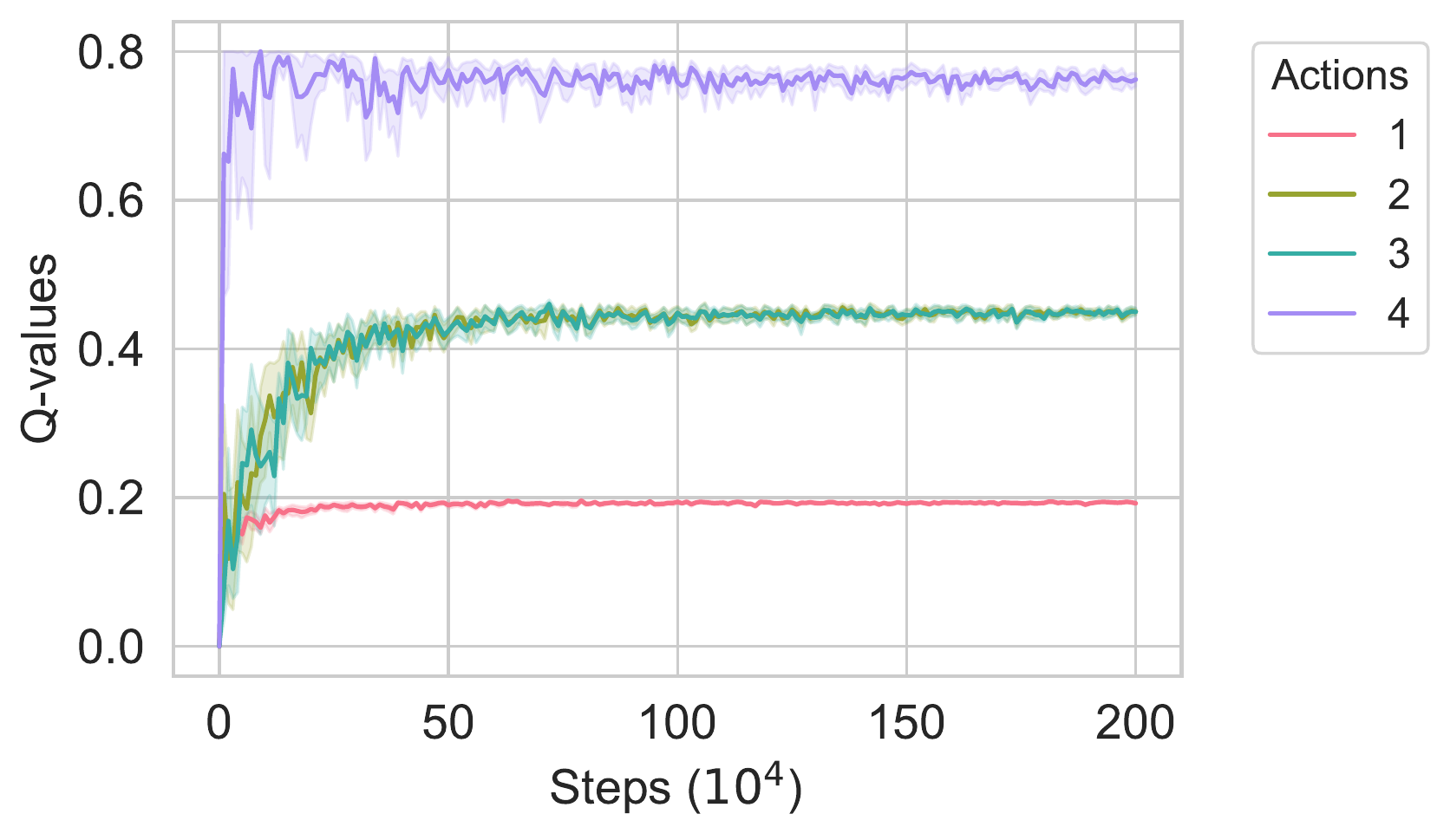}
		\subcaption{$Q$-values}\label{fig:QConverge}
	\end{minipage}%
	\begin{minipage}{0.49\textwidth}
		\centering
		\includegraphics[width=0.95\textwidth]{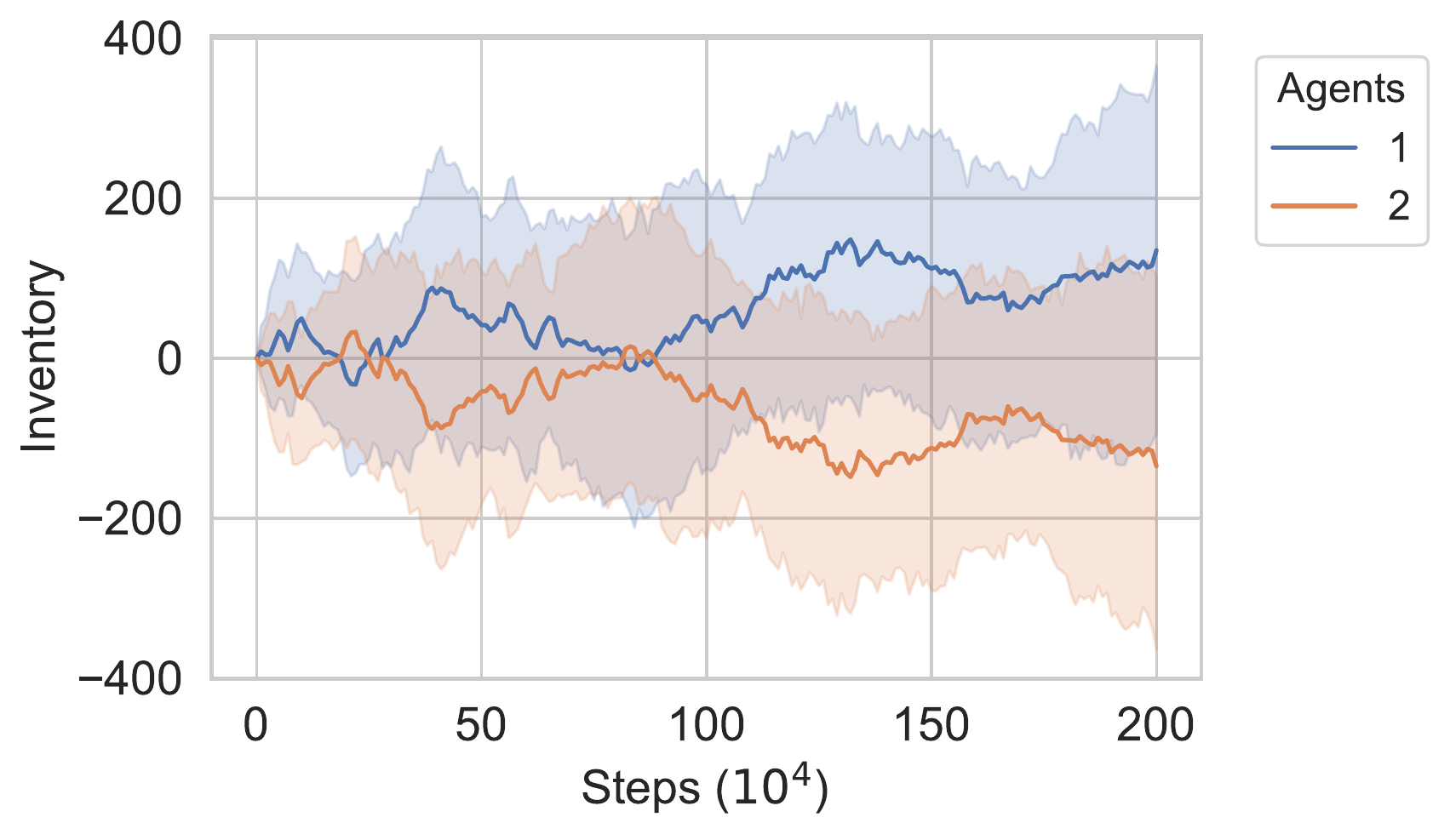}
		\subcaption{Inventory}\label{fig:Inventory}
	\end{minipage}
	\caption{Experiments under Table \ref{tab:SH}'s SH game with inventory risk. The left sub-graph shows the convergence of $Q$-values for each action. The right sub-graph illustrates the average inventory levels of each agent. Temperature $\lambda = 0.1$ and inventory risk aversion $\xi = 0.1$. Solid curves are mean values and shadowed areas are confidence intervals, both calculated with ten instances. Actions are encoded as in Table \ref{tab:inven}. Algorithms terminate after two million steps with convergence.}\label{fig:InvRisk}
\end{figure}
Experimentally, we observe that our algorithms have converged to the theoretical $Q$-values $q^*$, as illustrated in Figure \ref{fig:QConverge}. The algorithms start from zero $Q$-values and terminate after two million periods. Figure \ref{fig:Inventory} shows that market makers have small inventories during the whole training progression. 

\begin{figure}[H]
	\centering
	\begin{minipage}{0.49\textwidth}
		\centering
		\includegraphics[width=0.95\textwidth]{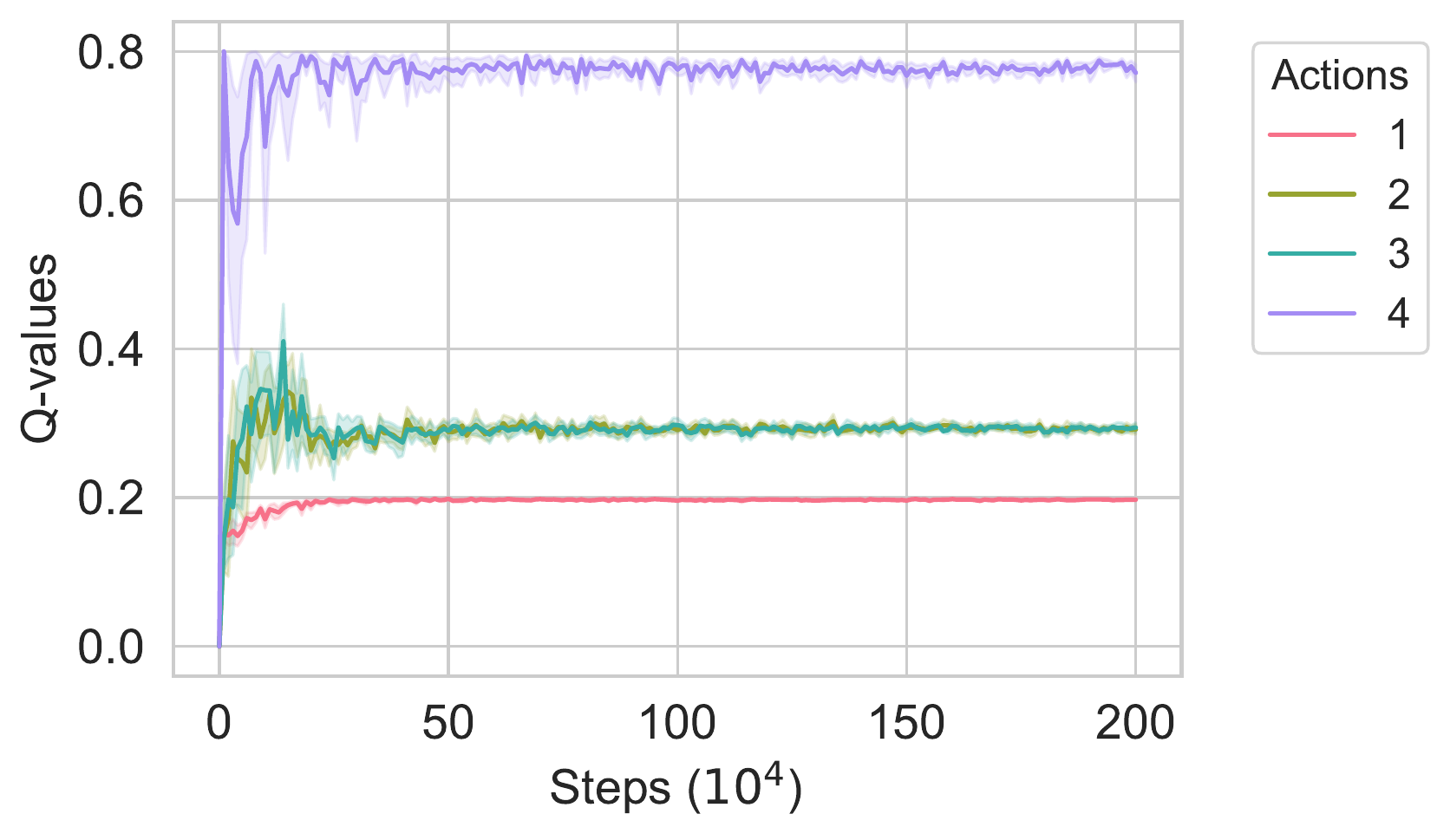}
		\subcaption{$Q$-values}\label{fig:QHard}
	\end{minipage}%
	\begin{minipage}{0.49\textwidth}
		\centering
		\includegraphics[width=0.95\textwidth]{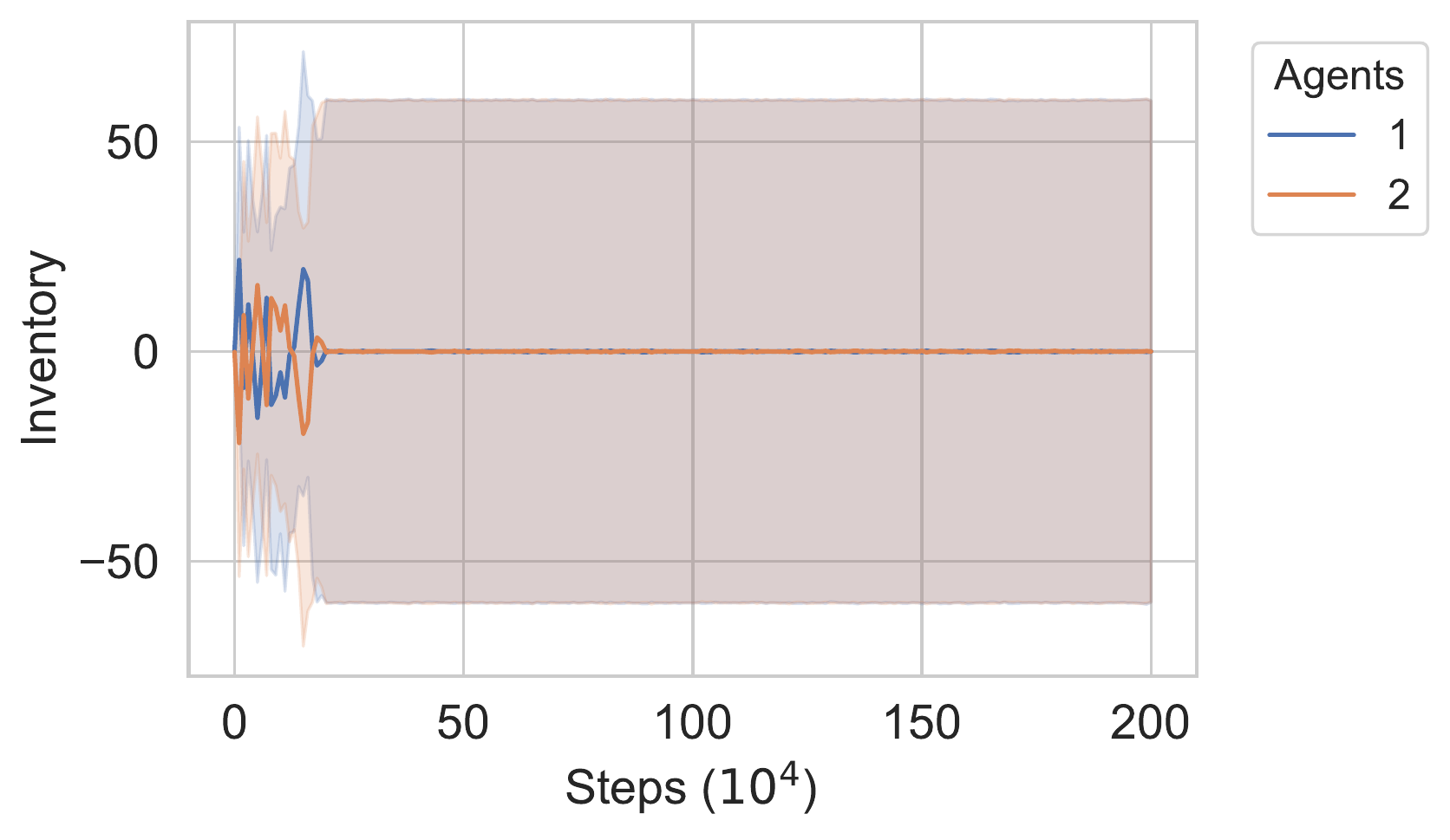}
		\subcaption{Inventory}\label{fig:InvHard}
	\end{minipage}
	\caption{Inventory risk management by skewing orders. If the inventory is higher than 100, then the agent sets ask$=0.1$ and bid$=0.8$. If it is lower than $-100$, set ask$=0.8$ and bid$=0.1$ instead. Let $\xi = 0$. Other parameters are the same as in Figure \ref{fig:InvRisk}.}\label{fig:Hard}
\end{figure}

Nevertheless, our formulation above uses an $L^2$ penalty on inventory changes. Market makers could also consider skewing its spreads \citep{nips19MM}. If the inventory rises to an upper bound, the agent can set the ask spread low and the bid spread high. Similarly, it can offset negative inventory by doing the opposite. In the experiment of Figure \ref{fig:Hard}, we suppose agents consider $-100$ and $100$ as thresholds. The skewed spreads are adopted once the inventory reaches the boundaries. Figure \ref{fig:InvHard} illustrates that the inventories are less volatile. Moreover, the confidence intervals are more restrictive than the thresholds, since they are averaged across instances. In a specific instance, the thresholds can be reached. The theoretical $Q$-values are not directly obtainable. Figure \ref{fig:QHard} shows lower $Q$-values for actions $(0.1, 0.8)$ and $(0.8, 0.1)$, compared with the $L^2$-penalty case. More importantly, the skewing method does not weaken the cooperation between agents.

\subsection{Multiple actions}\label{sec:multi}
Previous results reveal that algorithms with PD payoffs and two spreads converge to the Nash equilibrium, i.e., the lowest spread. In this subsection, we extend the results to multiple actions (spreads). Surprisingly, the lowest spread has relatively small probabilities under suitable temperature values.

\begin{figure}[H]
	\centering
	\includegraphics[width=0.6\linewidth]{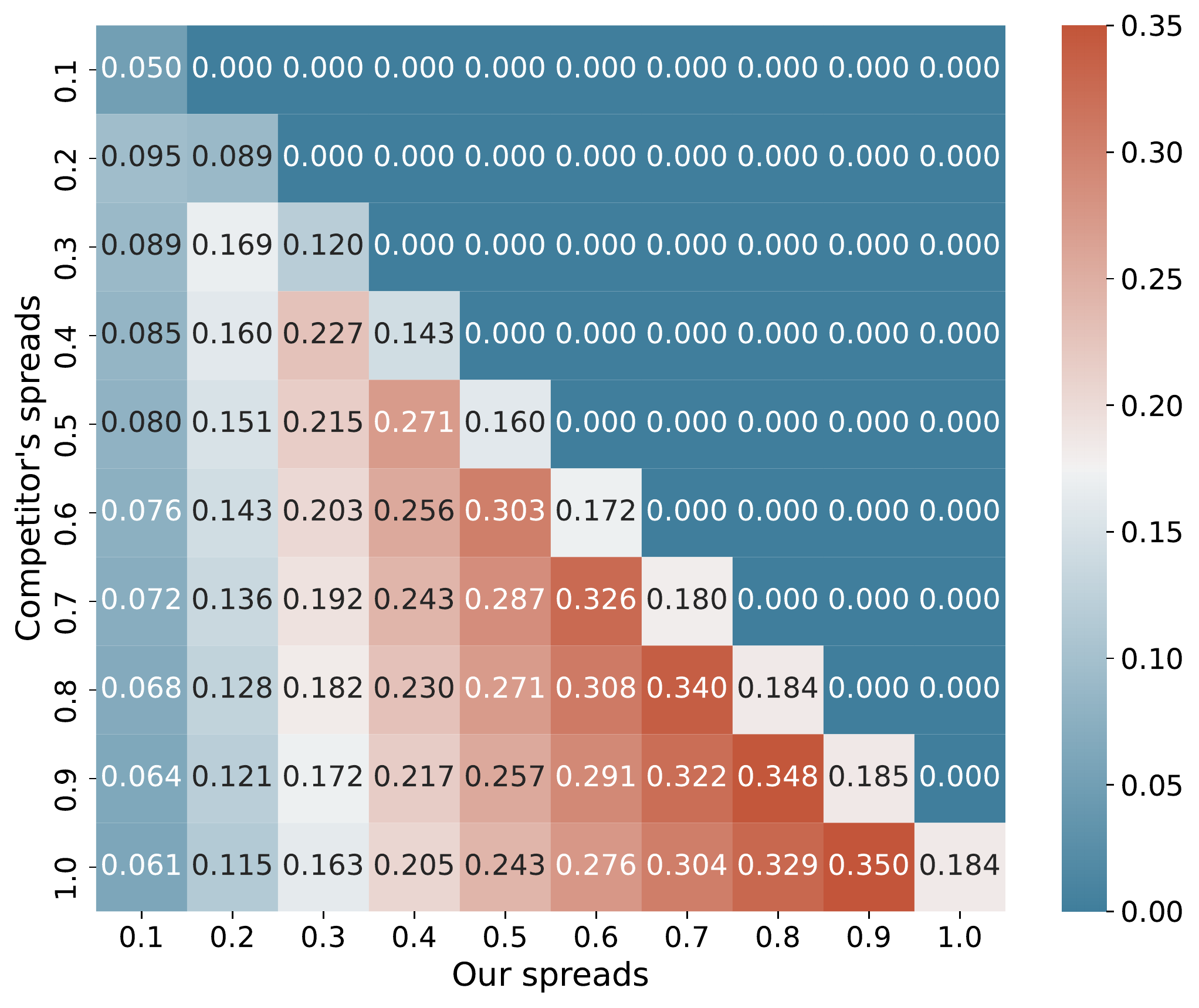}
	\caption{One-side payoff matrix when agents ignore inventory risk. Available spreads are evenly spaced as $\{0.1, 0.2, ..., 0.9, 1.0\}$. Weights $\{0, 1/90, ..., 9/90\}$. Volatility $\sigma = 0.1$. The horizontal axis represents our spreads. The competitor is on the vertical axis. Entries show the rewards in one period for ourselves.}
	\label{fig:oneside}
\end{figure}

\begin{table}
	\small
	\centering
	\begin{tabular}{c  c  c  c  c c c c c c c }
		\hline
		Spreads & 0.1 & 0.2 & 0.3 & 0.4 & 0.5 & 0.6 & 0.7 & 0.8 & 0.9 & 1.0 \\ 
		\hline
		$Q$-values & 0.0783 & 0.1270 & 0.1421 & 0.1324 & 0.1114 & 0.0876 & 0.0646 & 0.0436 & 0.0247 & 0.0080 \\
		\hline
		Probability & 0.0878 & 0.1429 & {\bf 0.1662} & 0.1509 &  0.1223 & 0.0964 & 0.0766 & 0.0621 & 0.0514 & 0.0435 \\
		\hline
	\end{tabular}
	\caption{Theoretical $Q$-values and probabilities in Figure \ref{fig:oneside} with temperature $\lambda = 0.1$.}	\label{tab:oneside}
\end{table}

\begin{figure}
	\centering
	\includegraphics[width=0.5\linewidth]{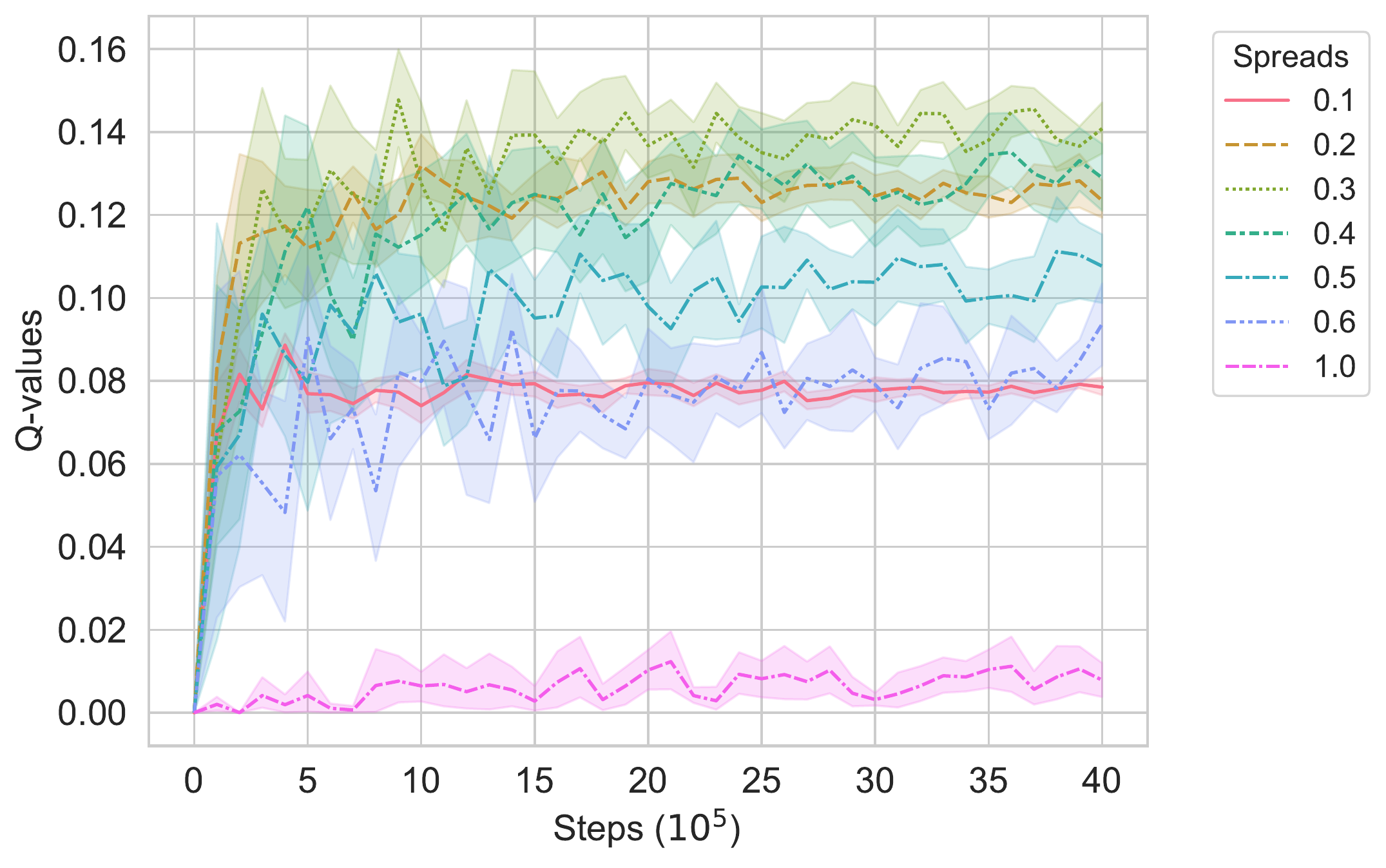}
	\caption{Converged $Q$-values with the payoff in Figure \ref{fig:oneside}. Temperature $\lambda = 0.1$. Mean values and confidence intervals are based on ten instances. Algorithms terminate after four million steps. Learning rate is set to $10^5/(10^5 + t)$. Only seven spreads are shown for readability.}
	\label{fig:Qoneside}
\end{figure}

In the first example, suppose the agents do not consider the inventory risk. There are ten spreads available for the ask (or bid) side, evenly spaced as $\{0.1, 0.2, ..., 0.9, 1.0\}$. Market orders arrive randomly. The weights for ten spreads are $\{0, 1/90, ..., 9/90\}$, respectively. The mid-price volatility is $0.1$. Therefore, when both agents select $0.1$, the market order arrives with probability one. The arrival probability is a decreasing function of the spread. If they both charge $1.0$, the probability is reduced to $\exp(-9/90/0.1) = 36.79\%$. Figure \ref{fig:oneside} illustrates the payoff for one agent. The transposed matrix is the payoff for another agent by symmetry. $(0.1, 0.1)$ is the unique Nash equilibrium under this setting. However, with a temperature $\lambda = 0.1$, the theoretical long-run probabilities and $Q$-values are given in Table \ref{tab:oneside}. Four spreads $\{0.2, 0.3, 0.4, 0.5\}$ contribute to 58.23\% of the charged spreads in the long-run. Spread $0.3$ has the highest probability and almost doubles the counterpart for spread $0.1$. Experimentally, the algorithm also converges to the theoretical $Q$-values and probabilities after short periods, as shown in Figure \ref{fig:Qoneside}.

\begin{table}
	\small
	\centering
	\begin{tabular}{c  c  c  c  c c c c c}
		\hline
		Actions & 1 & 2 & 3 & 4 & 5 & 6 & 7 & 8  \\ 
		\hline
		$Q$-values & 0.1063 & 0.1088 & 0.0665 & 0.0256 & 0.1088 & 0.1333 & 0.1049 & 0.0706   \\
		\hline
		Probability & 0.0803 & 0.0823 & 0.0539 & 0.0358 & 0.0823 & {\bf 0.1051} & 0.0791 & 0.0562 \\
		\hline
		\hline
		Actions & 9 & 10 & 11 & 12 & 13 & 14 & 15 & 16 \\ 
		\hline
		$Q$-values & 0.0665 & 0.1049 & 0.0859 & 0.0565 & 0.0256 & 0.0706 & 0.0565 & 0.0297 \\
		\hline
		Probability & 0.0539 & 0.0791 & 0.0654 & 0.0488 & 0.0358 & 0.0562 & 0.0488 & 0.0373 \\
		\hline 
	\end{tabular}
	\caption{Theoretical $Q$-values and probabilities in Figure \ref{fig:twosides} with temperature $\lambda = 0.1$. Action 6 or $(7/30, 7/30)$ has the highest probability.} \label{tab:twosides}
\end{table}

\begin{figure}[H]
	\centering
	\begin{minipage}{0.49\textwidth}
		\centering
		\includegraphics[width=0.95\textwidth]{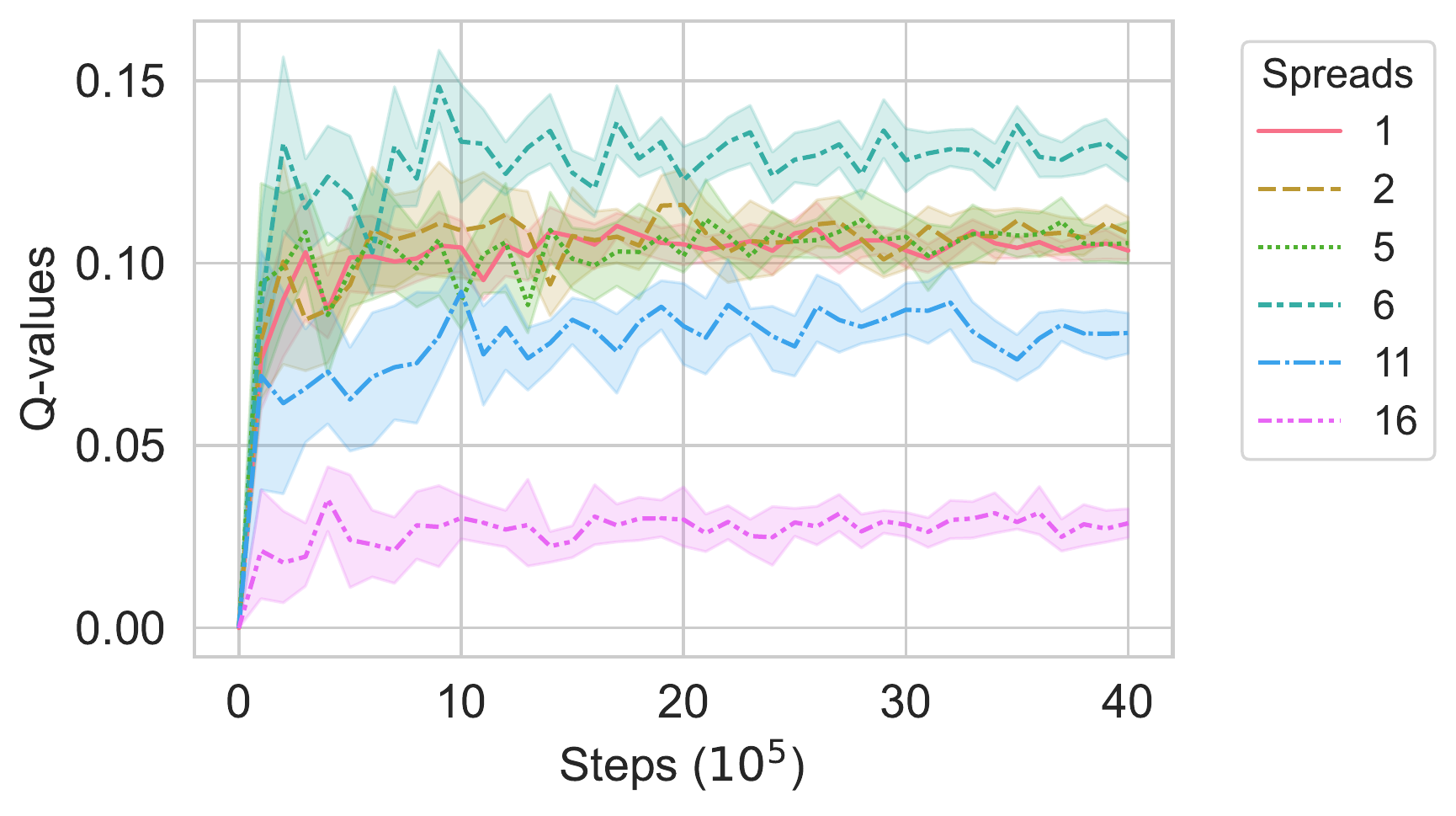}
		\subcaption{$Q$-values}\label{fig:Q_twosides}
	\end{minipage}%
	\begin{minipage}{0.49\textwidth}
		\centering
		\includegraphics[width=0.95\textwidth]{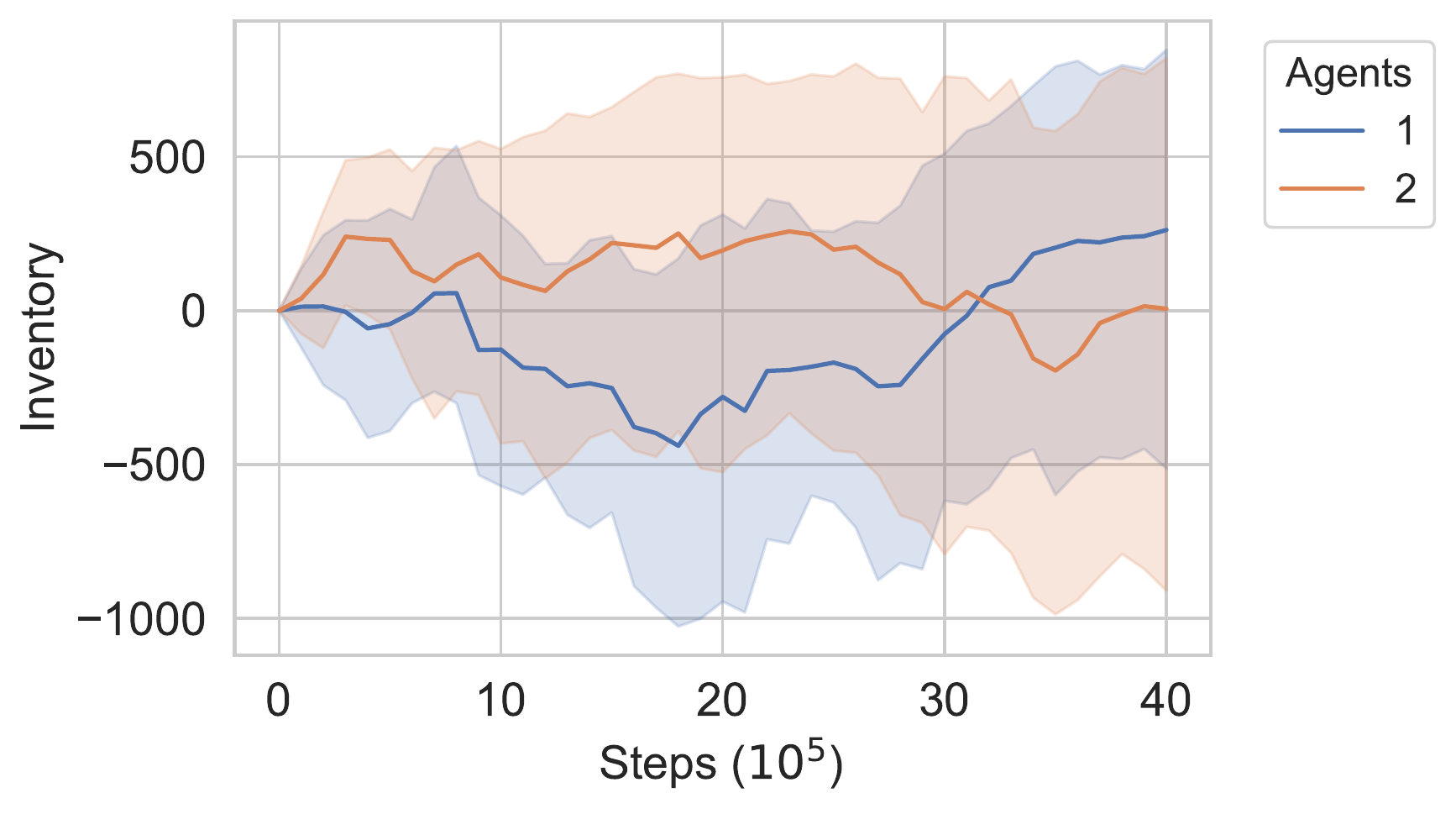}
		\subcaption{Inventory}\label{fig:Inv_twosides}
	\end{minipage}
	\caption{Experiments with the payoff in Figure \ref{fig:twosides}. Inventory risk aversion $\xi = 0.1$. Mean values and confidence intervals are calculated with ten instances. Algorithms terminate after four million steps. Learning rate is set to $10^5/(10^5 + t)$. Only six spreads are shown for readability.}\label{fig:ExpTwosides}
\end{figure}

The second example embraces the inventory risk by setting $\xi = 0.1$. Assume there are four spreads $\{3/30, 7/30, 11/30, 15/30\}$ on each side. The joint (ask, bid) actions are encoded with a row-major order. For example, the action 9 is referring to (ask = 11/30, bid = 3/30). The mid-price volatility is $0.1$. With weights $\{0, 1/30, 2/30, 3/30\}$ for these four spreads, the arrival probability is also decreasing in spreads. When both agents select the highest spread $0.5$, the probability of the market order in one side is also $\exp(-3/30/0.1) = 36.79\%$. Figure \ref{fig:twosides} in the Appendix gives the detailed payoff matrix. It shows the actions $1, 2, 5, 6$, or $(3/30, 3/30)$, $(3/30, 7/30)$, $(7/30, 3/30)$ and $(7/30, 7/30)$ are Nash equilibria. But with $\lambda = 0.1$, Table \ref{tab:twosides} indicates that $(7/30, 7/30)$ has the highest probability of 10.51\%, in contrast to 8.03\% for the lowest spread. Experiments in Figure \ref{fig:ExpTwosides} with $L^2$ inventory risk also demonstrate that the algorithm can converge to the theoretical $Q$-values. Figure \ref{fig:Inv_twosides} shows that the magnitude of inventory is larger since more choices on spreads induce higher fluctuations in the learning progression. The inventories are still low enough after millions of steps. Moreover, the temperature $\lambda$ has an essential effect on outcomes. If we set $\lambda = 0.01$, then the action 1 has the highest probability given by 99.62\%. Experiments also agree with the theoretical limit and are omitted here. While $\lambda = 0.01$ reduces the trading costs of investors, $\lambda = 0.1$ is disadvantageous to the investors. 

\subsection{Market making with memory}\label{sec:mem}
In general, Q-learning can incorporate states to reflect the current information known by agents \citep{watkins1992q,gordon2017,calvano2020artificial}. Suppose market makers observe the same discrete state variable $v(t)$, then the updating equation \eqref{Eq:update} is modified as
\begin{equation}\label{Eq:update-state}
	q_{i, t+1} (v(t), c_i(t)) = (1 - \alpha_t) q_{i, t} (v(t), c_i(t)) + \alpha_t [ r_i(t) + \gamma \max_{c'_i} q_{i,t}(v(t+1), c'_i)],
\end{equation}
where $v(t+1)$ is the next state. In this paper, we set the state $v(t)$ as the past action $C(t-1)$ by all agents. It implies that all agents can observe others' previous spreads and have a bounded memory in this section.

\begin{figure}
	\centering
	\includegraphics[width=0.5\linewidth]{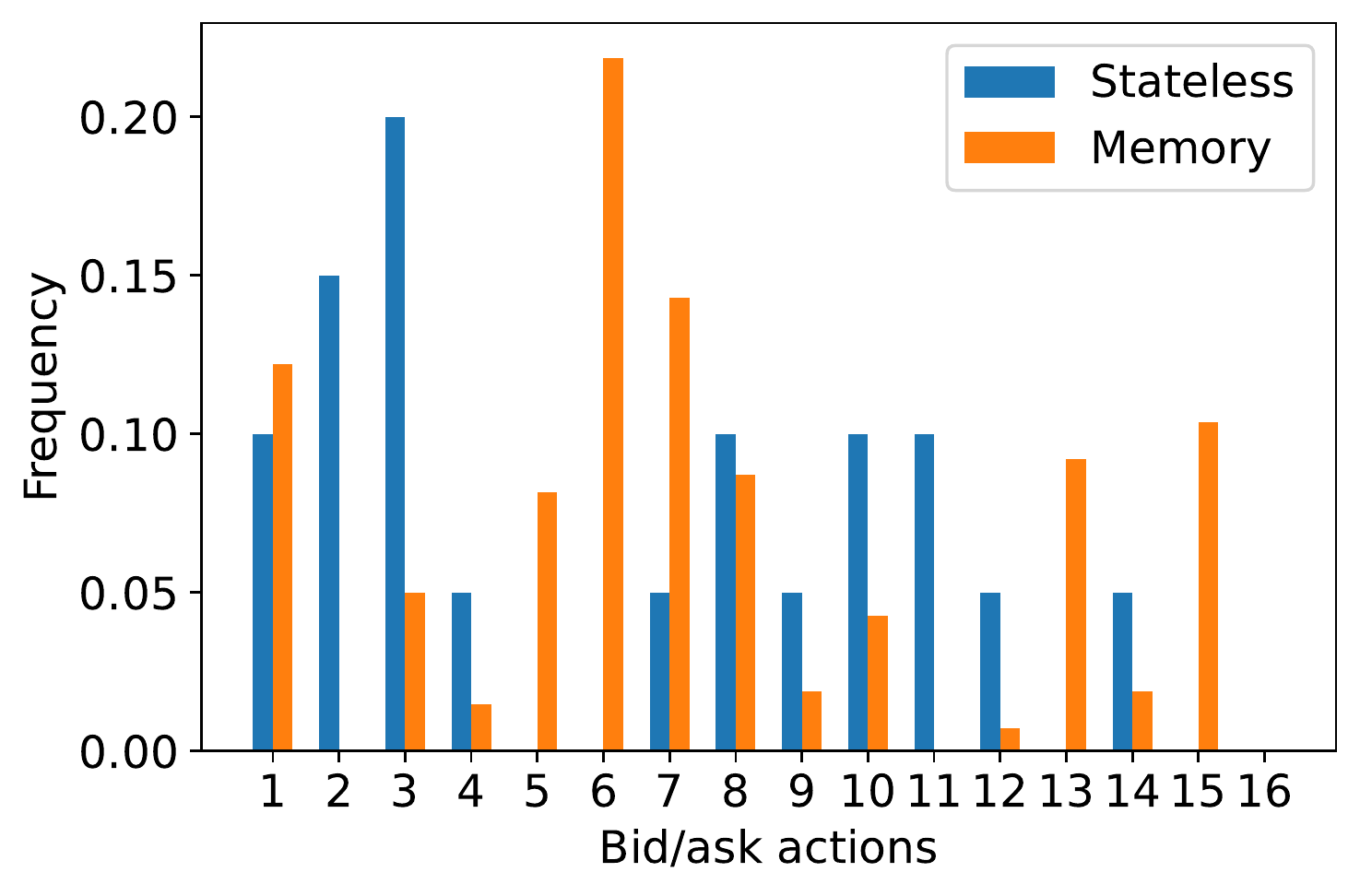}
	\caption{Empirical distributions of bid/ask actions with stateless Q-learning or memory. The same simulated data in Table \ref{tab:mem} are used. Some actions, such as 16, are never quoted.}
	\label{fig:mem}
\end{figure}

We still consider the payoff matrix in Figure \ref{fig:twosides} with multiple bid/ask spreads and inventory risk. In contrast to the previous temperature $\lambda = 0.1$ case, the action 1 has the probability with 99.62\% under stateless Q-learning when $\lambda = 0.01$. However, we observe that the agents still can cooperate and avoid this lowest bid/ask spread considerably when they have memory and a large discount factor $\gamma$. Table \ref{tab:mem} documents the average orders and rewards per period. The orders include both ask and bid sides. The rewards are for one agent and include the penalty of inventory. In the first panel with $\gamma=0$, myopic or short-sighted agents choose the lowest bid/ask spread, and thus the orders are close to two, under both the stateless Q-learning and the state-based setting. The difference appears when we consider far-sighted agents with $\gamma=0.9$. Rewards for agents with memory increase since they quote higher spreads more frequently. The number of orders reduces to 1.299 since investors are less willing to trade under high bid/ask spreads. Figure \ref{fig:mem} reveals more details of the distributions of the spreads with $\gamma=0.9$. Agents with memory quote the action 6 (ask=bid=7/30) with a frequency higher than 20\%. The stateless case favors the action 3 (ask=3/30, bid=11/30) instead. 

While the payoff matrix per period is unchanged, Table \ref{tab:mem} and Figure \ref{fig:mem} indicate that far-sighted agents with memory can avoid the lowest Nash equilibrium and favor a higher one with more profits. Nevertheless, the Q-learning with states is more volatile; see the standard deviations in Table \ref{tab:mem}. Unfortunately, the theoretical guarantee of these empirical observations is non-trivial and left as an open question.   

\begin{table}
	\centering
	\begin{tabular}{c | c c c}
		\hline
			  &    & stateless & memory \\ 
		\hline
		myopic & orders & 1.9967 (0.0591) & 1.9953 (0.0727)  \\
		($\gamma=0$) & rewards & 0.0996  (0.0063)& 0.0998 (0.0085)\\
		\hline
		far-sighted & orders & 1.3546 (0.6646) & 1.2990 (0.6768)  \\
		($\gamma=0.9$) & rewards & 0.0930 (0.0974) & 0.1123 (0.1192) \\
		\hline
	\end{tabular}
	\caption{Comparison of average orders and rewards when agents have memory. Standard deviations are in parentheses. Orders are for two agents and rewards are for one. Recall that there are at most two orders per period. We set $\lambda=0.01$ and other parameters are the same as Figure \ref{fig:twosides}. Data are from the last 1000 periods of each instance.} \label{tab:mem}
\end{table}

With a two-agent assumption, higher spreads or even cooperative strategies are achievable with independent learning algorithms. Communication between agents is not needed. The current antitrust law is ill-equipped to prevent this cooperation, which harms the interest of investors. A natural question is how to stop it. The common wisdom is that cooperating with more agents is harder. If we increase the number of market makers, can we guarantee the algorithms charge lower spreads with probability one after convergence? In the next section, we study the limiting behavior when the number of agents $N \to \infty$. 

\section{Infinite number of agents}\label{Sec:InfAgents}
The following decomposition result is for later use. Briefly speaking, if the rewards can be separated into ask and bid sides, then the market-making decision is also separable. 
\begin{lemma}\label{Lem:Sep}
	If the total reward is separable in the sense that
	\begin{equation}\label{eq:SepR}
	\E [r_i | C] =  \E[r_{i, a} |A] + \E[r_{i, b}| B] + r_d,
	\end{equation} 
	where $r_d$ is a deterministic constant independent of agent $i$ and action $C$. Then a solution $q^*(w)$ to \eqref{Eq:q*} can be decomposed into ask side, bid side, and a deterministic constant. That is,
	\begin{align}
	q^* (w) & = q^*_a (w_a) + q^*_b (w_b) + \frac{r_d}{1 - \gamma}, \quad \text{ with } w = (w_a, w_b),  \label{Eq:Separa} \\
	q^*_a(w_a) & = \sum_{A |_{a_1 = w_a}} \left\{ \Big[\prod_{j \neq 1} \cL (q^*_a, a_j) \Big] \E[r_{1,a} | A] \right\} + \gamma \max_{w'_a} q^*_a(w'_a),  \label{Eq:q*a} \\
	q^*_b(w_b) & = \sum_{B |_{b_1 = w_b}} \left\{ \Big[\prod_{j \neq 1} \cL (q^*_b, b_j) \Big] \E[r_{1,b} | B] \right\} + \gamma \max_{w'_b} q^*_b(w'_b).
	\end{align} 
\end{lemma}

Lemma \ref{Lem:Sep} agrees with our intuition. However, not every action selection mechanism implies the conclusion in Lemma \ref{Lem:Sep}. For example, another commonly used specification \citep{hart2000simple,unified} is 
\begin{equation*}
\frac{\max\{0, q_i(c_i)\}}{\sum_{c'} \max\{0, q_i(c')\}}.
\end{equation*}
It does not satisfy the separability as in \eqref{Eq:SepBoltz}. This is another motivation to consider Boltzmann selection \eqref{Eq:Boltz}.

Next, we study the limiting behavior when the number of agents goes to infinity. Note that for (i) and (iii) in Theorem \ref{Thm:Limit}, a specification of the arrival probability such as \eqref{Eq:ArrivalP} is not needed.
\begin{theorem}\label{Thm:Limit}
	Suppose for any number of agents, the order arrival probabilities $\cP(A)$ and $\cP(B)$ are lower bounded by a positive constant $0 < \varepsilon < 1$. $\xi/N \rightarrow 0$ and $\lambda \sim \frac{1}{N^u}$ with a constant $u$, as $N \rightarrow \infty$. Then market making on bid/ask sides is separable. Denote $x^* = (x^*_1, ..., x^*_M)$ as the corresponding probability of $M$ one-side ask (bid) spreads when $N \rightarrow \infty$. 
	\begin{enumerate}
		\item If $u < 1$, then $x^* = \id/M$ and each spread is selected equally;
		\item $u = 1$. For a given ask spreads vector $A$, recall the vector $v \in \R^M$ with $v_i$ as the number of agents choosing spread $i$. Suppose $\cP(A)$ depends on $A$ only through the frequency vector $v/N$. That is, $\cP(A) = \cP(v/N)$. Moreover, suppose $\cP(v/N)$ is continuous in $\R^M$. Then $x^*_1$, the probability of the lowest spread satisfies
		\begin{equation}
		x^*_1 = \frac{\exp( \frac{K(1)\cP(x^*)}{x^*_1})}{ \exp( \frac{K(1)\cP(x^*)}{x^*_1}) + M - 1}.
		\end{equation}
		$x^*_i = (1 - x^*_1)/(M-1)$ for all $i = 2, ... , M$.
		\item If $u > 1$, then $x^*_1 = 1$. Only the lowest spread is selected.
	\end{enumerate}
\end{theorem}

Theorem \ref{Thm:Limit} indicates that even with an infinite number of agents, they do not necessarily select the lowest spread with probability one. One reason is that all spreads become almost the same since they have to split the order with infinite competitors. The temperature parameter $\lambda$ then plays an essential role in distinguishing which spread is more profitable. Secondly, even when the theoretical limit selects the lowest spread, the algorithms do not necessarily converge with many agents. A careful investigation of the proof of Theorem \ref{Thm:Limit} shows that, for a sufficiently large $N$, 
\begin{equation*}
	\max_w \frac{|R_{1,a}(w; q^*_a)|}{\lambda} = \frac{|R_{1,a}(1; q^*_a)|}{\lambda} = O\left(\frac{1}{\lambda N} \right).
\end{equation*}
For Theorem \ref{Thm:Contraction}, we need $\lambda \sim 1/N^u$ with $u<0$ to guarantee the contraction property. The spreads are selected uniformly in this case. For other cases in Theorem \ref{Thm:Limit}, the convergence becomes unclear.

Given these results, including many market makers is not an efficient way to prevent cooperative outcomes.

\section{Discussions and future works}\label{Sec:Dis}
This paper formulates the market-making competition as a repeated matrix game. The advantage of our framework is a multi-agent setting and a flexible relationship between market order arrivals and market liquidity. Inventory risk appears as a cost in the payoff. 

A natural question is if cooperation with independent learning is possible, why have we not observed it yet in the real market? There are several explanations. First, investors may be sensitive to the changes in spreads, such that a prisoner's dilemma payoff like Table \ref{tab:PD} is plausible. Then our algorithms also converge to the lower spread. Second, market makers are very likely to use heterogeneous algorithms. Third, restrictions in our market modeling may facilitate cooperation. Generalizations on our model may reveal new factors that are crucial for cooperation. For example, one can consider agents acting asynchronously in continuous time. The effect of adverse selection and informed investors is also a potential direction. Besides, a long-standing open problem is the convergence analysis under Q-learning with multi-states and multi-agents.   

From the regulatory perspective, no antitrust law has been created to targeting tacit coordination between autonomous algorithms. One reason is that this form of coordination has not been observed yet in the real market. It becomes unjustified to create laws for these unobserved conducts. Nevertheless, it is crucial to realize that cooperation with independent learning is not completely impossible. Market participants and regulators may anticipate what kind of regulations could be created if this form of coordination becomes a market reality. 

\section*{Acknowledgments}
Bingyan Han is supported by UIC Start-up Research Fund (Reference No: R72021109). The author would like to thank the anonymous referees and the editors for their careful reading and valuable comments, which have greatly improved the manuscript. Section \ref{sec:mem} is motivated by a suggestion from the referee.


\begin{thebibliography}{}
	
	\bibitem[Avellaneda and Stoikov, 2008]{avellaneda2008high}
	Avellaneda, M. and Stoikov, S. (2008).
	\newblock High-frequency trading in a limit order book.
	\newblock {\em Quantitative Finance}, 8(3):217--224.
	
	\bibitem[Baldacci et~al., 2021]{baldacci2021optimal}
	Baldacci, B., Possama{\"\i}, D., and Rosenbaum, M. (2021).
	\newblock Optimal make-take fees in a multi market-maker environment.
	\newblock {\em SIAM Journal on Financial Mathematics}, 12(1):446--486.
	
	\bibitem[Calvano et~al., 2020a]{calvano2020protecting}
	Calvano, E., Calzolari, G., Denicol{\`o}, V., Harrington, J.~E., and
	Pastorello, S. (2020a).
	\newblock Protecting consumers from collusive prices due to {AI}.
	\newblock {\em Science}, 370(6520):1040--1042.
	
	\bibitem[Calvano et~al., 2020b]{calvano2020artificial}
	Calvano, E., Calzolari, G., Denicol{\`o}, V., and Pastorello, S. (2020b).
	\newblock Artificial intelligence, algorithmic pricing, and collusion.
	\newblock {\em American Economic Review}, 110(10):3267--97.
	
	\bibitem[Chao and Strawderman, 1972]{chao1972negative}
	Chao, M.-T. and Strawderman, W. (1972).
	\newblock Negative moments of positive random variables.
	\newblock {\em Journal of the American Statistical Association},
	67(338):429--431.
	
	\bibitem[Christie et~al., 1994]{christie1994did}
	Christie, W.~G., Harris, J.~H., and Schultz, P.~H. (1994).
	\newblock Why did {NASDAQ} market makers stop avoiding odd-eighth quotes?
	\newblock {\em The Journal of Finance}, 49(5):1841--1860.
	
	\bibitem[Christie and Schultz, 1994]{christie1994nasdaq}
	Christie, W.~G. and Schultz, P.~H. (1994).
	\newblock Why do {NASDAQ} market makers avoid odd-eighth quotes?
	\newblock {\em The Journal of Finance}, 49(5):1813--1840.
	
	\bibitem[Christie and Schultz, 1995]{christie1995policy}
	Christie, W.~G. and Schultz, P.~H. (1995).
	\newblock Policy watch: Did {NASDAQ} market makers implicitly collude?
	\newblock {\em Journal of Economic Perspectives}, 9(3):199--208.
	
	\bibitem[Claus and Boutilier, 1998]{claus1998dynamics}
	Claus, C. and Boutilier, C. (1998).
	\newblock The dynamics of reinforcement learning in cooperative multiagent
	systems.
	\newblock {\em AAAI/IAAI}, (746-752).
	
	\bibitem[Ezrachi and Stucke, 2016]{ezrachi2016virtual}
	Ezrachi, A. and Stucke, M.~E. (2016).
	\newblock {\em Virtual competition}.
	\newblock Oxford University Press.
	
	\bibitem[Foupouagnigni and Mouafo~Wouodji{\'e},
	2020]{foupouagnigni2020multivariate}
	Foupouagnigni, M. and Mouafo~Wouodji{\'e}, M. (2020).
	\newblock On multivariate {B}ernstein polynomials.
	\newblock {\em Mathematics}, 8(9):1397.
	
	\bibitem[Ganesh et~al., 2019]{nips19MM}
	Ganesh, S., Vadori, N., Xu, M., Zheng, H., Reddy, P.~P., and Veloso, M. (2019).
	\newblock Reinforcement learning for market making in a multi-agent dealer
	market.
	\newblock In {\em Advances in Neural Information Processing Systems}.
	
	\bibitem[Gordon, 2017]{gordon2017}
	Gordon, R. (2017).
	\newblock Machine learning for trading.
	\newblock {\em Risk}, 30(10):84--89.
	
	\bibitem[Hansen et~al., 2020]{hansen2020algorithmic}
	Hansen, K., Misra, K., and Pai, M. (2020).
	\newblock Algorithmic collusion: Supra-competitive prices via independent
	algorithms.
	\newblock {\em Marketing Science}.
	
	\bibitem[Hart and Mas-Colell, 2000]{hart2000simple}
	Hart, S. and Mas-Colell, A. (2000).
	\newblock A simple adaptive procedure leading to correlated equilibrium.
	\newblock {\em Econometrica}, 68(5):1127--1150.
	
	\bibitem[Jaakkola et~al., 1994]{jaakkola1994convergence}
	Jaakkola, T., Jordan, M.~I., and Singh, S.~P. (1994).
	\newblock On the convergence of stochastic iterative dynamic programming
	algorithms.
	\newblock {\em Neural computation}, 6(6):1185--1201.
	
	\bibitem[Klein, 2019]{klein2019autonomous}
	Klein, T. (2019).
	\newblock Autonomous algorithmic collusion: Q-learning under sequential
	pricing.
	\newblock {\em Working Paper}.
	\newblock \url{https://ssrn.com/abstract=3195812}.
	
	\bibitem[Lanctot et~al., 2017]{unified}
	Lanctot, M., Zambaldi, V.~F., Gruslys, A., Lazaridou, A., Tuyls, K.,
	P{\'{e}}rolat, J., Silver, D., and Graepel, T. (2017).
	\newblock A unified game-theoretic approach to multiagent reinforcement
	learning.
	\newblock In {\em Advances in Neural Information Processing Systems 30}, pages
	4190--4203.
	
	\bibitem[Lindsey et~al., 2016]{lindsey2016sec}
	Lindsey, R., Byrne, J.~A., and Schwartz, R.~A. (2016).
	\newblock The sec's order handling rules of 1997 and beyond: Perspective and
	outcomes of the landmark regulation.
	\newblock {\em The Journal of Portfolio Management}, 42(3):56--64.
	
	\bibitem[Now{\'{e}} et~al., 2012]{Nowe12}
	Now{\'{e}}, A., Vrancx, P., and Hauwere, Y.~D. (2012).
	\newblock Game theory and multi-agent reinforcement learning.
	\newblock In Wiering, M.~A. and van Otterlo, M., editors, {\em Reinforcement
		Learning}, volume~12 of {\em Adaptation, Learning, and Optimization}, pages
	441--470. Springer.
	
	\bibitem[Waltman and Kaymak, 2008]{waltman2008q}
	Waltman, L. and Kaymak, U. (2008).
	\newblock Q-learning agents in a {Cournot} oligopoly model.
	\newblock {\em Journal of Economic Dynamics and Control}, 32(10):3275--3293.
	
	\bibitem[Watkins and Dayan, 1992]{watkins1992q}
	Watkins, C.~J. and Dayan, P. (1992).
	\newblock Q-learning.
	\newblock {\em Machine Learning}, 8(3-4):279--292.
	
	\bibitem[Wunder et~al., 2010]{Wunder10}
	Wunder, M., Littman, M.~L., and Babes, M. (2010).
	\newblock Classes of multiagent {Q}-learning dynamics with $\varepsilon$-greedy
	exploration.
	\newblock In {\em International Conference on Machine Learning}, pages
	1167--1174. Omnipress.
	
\end{thebibliography}

\appendix
\section{Proofs of results}
\subsection{Proof of Lemma \ref{Lem:Contract}}
\begin{proof}
	We first consider the dependence of $\cH [q_i(w)]$ on other agents' $Q$-values. For $k \neq i$ and action index $l \in \{1, ..., M^2\}$,
	\begin{align*}
		\frac{\partial \cH [q_i(w)]}{\partial q_k(l)} =  \sum_{C |_{c_i = w}} \left\{  \E[r_i | C] \Big[\prod_{ j \neq i, k} \cL (q_j, c_j) \Big] \frac{\partial \cL(q_k, c_k)}{\partial q_k(l)}\right\}.
	\end{align*}
	If $c_k = l$, then 
	\begin{align*}
		\frac{\partial \cL(q_k, c_k)}{\partial q_k(l)} = \partial_{q_k(l)} \Big[ \frac{e^{q_k(l)/\lambda}}{\sum_{c'}  e^{q_k(c')/\lambda} } \Big] = \partial_{q_k(l)} \Big[ \frac{e^{q_k(l)/\lambda}}{e^{q_k(l)/\lambda} + E_c} \Big], 
	\end{align*}
	where $E_c$ is the sum of exponentials independent of $q_k(l)$. Then
	\begin{align*}
		\partial_{q_k(l)} \Big[ \frac{e^{q_k(l)/\lambda}}{e^{q_k(l)/\lambda} + E_c} \Big] = \frac{1}{\lambda} \frac{ e^{q_k(l)/\lambda} E_c}{(e^{q_k(l)/\lambda} + E_c)^2} = \frac{1}{\lambda} \cL(q_k, c_k) \big(1 - \cL(q_k, c_k)\big).
	\end{align*}
	Similarly, if $c_k \neq l$, then
	\begin{equation*}
		\frac{\partial \cL(q_k, c_k)}{\partial q_k(l)} = - \frac{1}{\lambda} \cL(q_k, c_k) \cL(q_k, l).
	\end{equation*}
	Overall, since $0 \leq \cL(q_k, \cdot) \leq 1$ is a probability,
	\begin{align}
		\left| \frac{\partial \cH[q_i(w)]}{\partial q_k(l)} \right| \leq \frac{1}{\lambda} |R_i(w; Q_{-i})|.
	\end{align}
	
	The dependence of $\cH[q_i(w)]$ on $q_i$ is via the maximal operator. Apply the mean value theorem to the first part of the $NM^2$-dimensional $\cH[Q] - \cH[Q']$, we obtain the claim in \eqref{Eq:Contract} as desired.
\end{proof}

\subsection{Proof of Theorem \ref{Thm:Contraction}}
\begin{proof}
	Subtracting $q^*(c_i(t))$ from both sides of the updating equation \eqref{Eq:update} and define
	\begin{align}
		\delta_{i, t}(c_i(t)) &:= q_{i, t}(c_i(t)) - q^*(c_i(t)), \\
		F_{i, t}(c_i(t)) &:= r_i (t) + \gamma \max_{c'_i} q_{i,t}(c'_i) - q^*(c_i(t)).
	\end{align}
	Then we obtain
	\begin{equation}
		\delta_{i, t+1}(c_i(t)) =  (1 - \alpha_t) \delta_{i, t}(c_i(t)) + \alpha_t F_{i, t}(c_i(t)).
	\end{equation}
	We verify that it satisfies the two conditions in \citet[Theorem 1]{jaakkola1994convergence}. First,
	\begin{align*}
		\E[ F_{i, t}(w) | \cF_t] & = \sum_{C |_{c_i = w}} \left\{ \Big[\prod_{j \neq i} \cL (q_{j,t}, c_j) \Big] \E[r_i(t) | C] \right\} + \gamma \max_{w'} q_{i,t}(w') - q^*(w) \\
		& = \cH[q_{i, t}(w)] - q^*(w).
	\end{align*}
	
	Since $ \cH[q^*(w)] = q^*(w)$, we have $\E[ F_{i, t}(w) | \cF_t] =  \cH[q_{i, t}(w)] - \cH[q^*(w)]$. By the contraction property \eqref{Eq:Contract}, 
	\begin{equation}
		|| \E[ F_{i, t}(w) | \cF_t] ||_\infty \leq d || Q_t - Q^*||_\infty = d || \Delta_t ||_\infty,
	\end{equation}
	where $\Delta_t$ is the vector of all $\delta_{i, t}$.
	
	Second,
	\begin{align*}
		\text{Var} [ F_{i, t}(w) | \cF_t] &= \E\Big[ \big(r_i (t) + \gamma \max_{c'_i} q_{i,t}(c'_i) - q^*(w) - \cH[q_{i, t}(w)] + \cH[q^*(w)] \big)^2 \Big] \\
		&= \E\Big[ \big(r_i(t) + \gamma \max_{c'_i} q_{i,t}(c'_i) - \cH[q_{i, t}(w)] \big)^2 \Big] \\
		& = \text{Var} \Big[ r_i(t) + \gamma \max_{c'_i} q_{i,t}(c'_i) | \cF_t \Big].
	\end{align*}
	Since $r_i$ is bounded, then
	\begin{equation*}
		\text{Var} [ F_{i, t}(w) | \cF_t] \leq \kappa (1 + || \Delta_t ||^2_W),
	\end{equation*}
	for some constant $\kappa$ and weighted maximal norm $ || \cdot ||_W$ \citep{jaakkola1994convergence}.
	
	Therefore, $\Delta_t$ converges to zero with probability one by \citet[Theorem 1]{jaakkola1994convergence}, which is exactly $q_{i,t}(w) \to q^*(w)$.
\end{proof}

\subsection{Proof of Lemma \ref{Lem:Sep}}
\begin{proof}
	We only need to verify that \eqref{Eq:Separa} is a solution to \eqref{Eq:q*}. With \eqref{Eq:Separa}, the Boltzmann selection becomes 
	\begin{align}
		\cL(q^*, c_j) &= \frac{e^{q^*(c_j)/\lambda}}{\sum_{c'}  e^{q^*(c')/\lambda}} = \frac{e^{ \frac{ q^*_a(a_j) + q^*_b(b_j) + r_d/(1-\gamma) }{ \lambda }}}{\sum_{c'} e^{ \frac{ q^*_a(a') + q^*_b(b') + r_d/(1-\gamma)}{ \lambda }}} \nonumber \\
		& = \frac{e^{ \frac{ q^*_a(a_j) }{ \lambda }}}{\sum_{a'} e^{ \frac{ q^*_a(a')}{ \lambda }}} \frac{e^{ \frac{ q^*_b(b_j) }{ \lambda }}}{\sum_{b'} e^{ \frac{ q^*_b(b')}{ \lambda }}} = \cL(q^*_a, a_j) \cL(q^*_b, b_j). \label{Eq:SepBoltz}
	\end{align}
	The expected values satisfy \eqref{eq:SepR} and then
	\begin{align*}
		& \sum_{C |_{c_1 = w}} \left\{ \Big[\prod_{j \neq 1} \cL (q^*, c_j) \Big] \E[r_1 | C] \right\} = \sum_{C |_{c_1 = w}} \left\{ \Big[\prod_{j \neq 1} \cL (q^*, c_j) \Big] (\E[r_{1,a} | A] + \E[r_{1,b} | B] + r_d) \right\}.
	\end{align*}
	We deal with these three parts individually. Note
	\begin{align*}
		\sum_{C |_{c_1 = w}} \left\{ \Big[\prod_{j \neq 1} \cL (q^*, c_j) \Big] \E[r_{1,a} | A] \right\} & =  \sum_{C |_{c_1 = w}} \left\{ \Big[\prod_{j \neq 1} \cL(q^*_a, a_j) \cL(q^*_b, b_j) \Big] \E[r_{1,a} | A] \right\}  \\
		& = \sum_{A|_{a_1 = w_a}} \left\{ \Big[\prod_{j \neq 1} \cL(q^*_a, a_j) \sum_{B | b_1 = w_b} \prod_{ j \neq 1} \cL(q^*_b, b_j) \Big] \E[r_{1,a} | A] \right\} \\
		& = \sum_{A|_{a_1 = w_a}} \left\{ \Big[\prod_{j \neq 1} \cL(q^*_a, a_j) \Big] \E[r_{1,a} | A] \right\},
	\end{align*}
	where we have used the equity $\sum_{B | b_1 = w_b} \prod_{ j \neq 1} \cL(q^*_b, b_j) = 1$. Similarly, we can calculate the bid side. Since $r_d$ is deterministic and independent of actions $C$, then $\sum_{C | c_1 = w} \Big[ \prod_{ j \neq 1} \cL(q^*, c_j) \Big] r_d = r_d$.
	
	Combine the derivations above, we obtain
	\begin{align*}
		& \sum_{C |_{c_1 = w}} \left\{ \Big[\prod_{j \neq 1} \cL (q^*, c_j) \Big] \E[r_1 | C] \right\} \\
		& = \sum_{A|_{a_1 = w_a}} \left\{ \Big[\prod_{j \neq 1} \cL(q^*_a, a_j) \Big] \E[r_{1,a} | A] \right\} + \sum_{B|_{b_1 = w_b}} \left\{ \Big[\prod_{j \neq 1} \cL(q^*_b, b_j) \Big] \E[r_{1,b} | B] \right\} + r_d.
	\end{align*}
	
	Since $w$ enumerates all possible combinations of $w_a$ and $w_b$, then
	\begin{align*}
		\max_{w'} q^*(w') = \max_{w'_a} q^*_a(w'_a) + \max_{w'_b} q^*_b(w'_b) + \frac{r_d}{1 - \gamma}.
	\end{align*}
	Finally, the right-hand side of \eqref{Eq:q*} reduces to
	\begin{align*}
		& \sum_{C |_{c_1 = w}} \left\{ \Big[\prod_{j \neq 1} \cL (q^*, c_j) \Big] \E[r_1 | C] \right\} + \gamma \max_{w'} q^*(w') \\
		= & \sum_{A|_{a_1 = w_a}} \left\{ \Big[\prod_{j \neq 1} \cL(q^*_a, a_j) \Big] \E[r_{1,a} | A] \right\} + \sum_{B|_{b_1 = w_b}} \left\{ \Big[\prod_{j \neq 1} \cL(q^*_b, b_j) \Big] \E[r_{1,b} | B] \right\} \\
		& + r_d + \gamma \max_{w'_a} q^*_a(w'_a) + \gamma \max_{w'_b} q^*_b(w'_b) + \gamma \frac{r_d}{1 - \gamma} \\
		= & \sum_{A|_{a_1 = w_a}} \left\{ \Big[\prod_{j \neq 1} \cL(q^*_a, a_j) \Big] \E[r_{1,a} | A] \right\} + \gamma \max_{w'_a} q^*_a(w'_a)  \\
		& + \sum_{B|_{b_1 = w_b}} \left\{ \Big[\prod_{j \neq 1} \cL(q^*_b, b_j) \Big] \E[r_{1,b} | B] \right\} + \gamma \max_{w'_b} q^*_b(w'_b)\\
		& + \frac{r_d}{1 - \gamma}. 
	\end{align*}
	It is exactly $q^*(w)$, which validates that \eqref{Eq:Separa} is a solution to \eqref{Eq:q*}.
\end{proof}

\subsection{Proof of Theorem \ref{Thm:Limit}}
\begin{proof}
	The total reward for agent 1 during one period satisfies the following inequality
	\begin{align*}
		r_{1,a} + r_{1,b} - \xi \leq r_1 \leq r_{1,a} + r_{1,b}.
	\end{align*}
	By Lemma \ref{Lem:Sep}, with either the upper or lower bounds above as the payoffs, the corresponding $Q$-values are decomposable. Moreover, the upper and lower bounds induce the same $Q$-values, if $\xi/N \rightarrow 0$ as $N \rightarrow \infty$. We can focus on the ask side since the bid side is the same by symmetry. Denote $x_{N} = (x_{1,N}, ..., x_{M, N})$ as the probability of $M$ ask spreads. The subscripts highlight that there are $N$ agents. 
	
	By the design of the payoff, reward $r_{1, a}$ is non-zero under $w_a$ only when all other agents quote spreads not lower than $K(w_a)$. Divide both sides of \eqref{Eq:q*a} by $\lambda$, we consider the limit of 
	\begin{equation}
		\sum_{A |_{a_1 = w_a}} \left\{ \Big[\prod_{j \neq 1} \cL (q^*_a, a_j) \Big] \frac{\E[r_{1,a} | A]}{\lambda} \right\} =: \frac{R_{1, a}(w_a; q^*_a)}{\lambda}.
	\end{equation}
	
	Start from the highest spread available. $\E[r_{1,a} | a_1 = M]$ is non-zero only when others select $a_j = M$. Since the arrival probability is lower bounded by $\varepsilon$, then 
	\begin{equation}\label{Eq:BoundM}
		\varepsilon x^{N-1}_{M,N} \frac{K(M)}{N\lambda} \leq \frac{R_{1, a}(M; q^*_a)}{\lambda} \leq x^{N-1}_{M, N} \frac{K(M)}{N\lambda},
	\end{equation}
	where $x_{M, N}$ is the probability of action $M$ when there are $N$ agents. $K(M)$ is the $M$-th spread.
	
	Similarly, for the second highest spread $M-1$,
	\begin{align}
		\frac{R_{1, a}(M - 1; q^*_a)}{\lambda} & \geq \varepsilon \binom{N-1}{k} x^{k}_{M-1, N} x^{N-1-k}_{M, N} \frac{K(M-1)}{k+1} \frac{1}{\lambda} \nonumber \\
		& = \frac{\varepsilon}{\lambda} (x_{M-1, N} + x_{M, N})^{N-1} \binom{N-1}{k} p^{k} (1 - p)^{N-1-k} \frac{K(M-1)}{k+1}  \nonumber \\
		& = \frac{\varepsilon}{\lambda} (x_{M-1,N} + x_{M,N})^{N-1} \E[ \frac{1}{Z + 1}] K(M-1), \label{Eq:BoundM1}
	\end{align}
	where $p = x_{M-1, N}/(x_{M-1, N} + x_{M, N})$ and $Z \sim \text{Binomial}(N-1, p)$. \cite{chao1972negative} proved that
	\begin{equation}
		\E[ \frac{1}{Z + 1}] = \frac{1 - (1-p)^N}{Np}.
	\end{equation}
	If for all $N$, $x_{M, N} \leq 1 - \tau$ with some $\tau > 0$, then $\frac{R_{1, a}(M; q^*_a)}{\lambda} \rightarrow 0$ for any polynomial order $u$, when $N \rightarrow \infty$. We argue that there is no subsequence of $x_{M, N}$ that converges to 1. Without loss of generality, denote the subsequence still as $x_{M, N}$. If $x_{M, N} \rightarrow 1$, then $x_{M-1, N} \rightarrow 0$ and $ p \rightarrow 0$. $\E[ \frac{1}{Z + 1}] $ converges to 1. Compare \eqref{Eq:BoundM} with \eqref{Eq:BoundM1}, $\frac{R_{1, a}(M-1; q^*_a)}{\lambda}$ is larger than $\frac{R_{1, a}(M; q^*_a)}{\lambda}$ when $N \to \infty$. It indicates that probability $x_{M-1, N}$ should be larger than $x_{M, N}$, which contradicts with the assumption that $x_{M, N} \rightarrow 1$. We can continue the argument to show that $\frac{R_{1, a}(w_a; q^*_a)}{\lambda} \rightarrow 0$ for any $w_a \geq 2$ and $\lambda \sim \frac{1}{N^u}$. Besides, we also obtain that $\sum^M_{i=2} x_{i, N}$ does not converge to 1. Therefore, $x_{1, N}$ is lower bounded by a positive constant.
	
	For the lowest spread, 
	\begin{align*}
		\frac{R_{1, a}(1; q^*_a)}{\lambda} \leq \frac{1}{\lambda} \frac{1 - (1 - x_{1, N})^N}{N x_{1, N}} K(1),
	\end{align*}
	which converges to 0 if $\lambda \sim \frac{1}{N^u}$ with $u < 1$. Then all $q^*_a(w)/\lambda$ go to zero, implying an equal probability for all spreads.
	
	When $u > 1$,
	\begin{align*}
		\frac{R_{1, a}(1; q^*_a)}{\lambda} \geq \frac{\varepsilon}{\lambda} \frac{1 - (1- x_{1, N} )^N}{N x_{1, N}} K(1) \rightarrow \infty.
	\end{align*}
	Only the lowest spread is selected when there are infinite agents.
	
	When $u = 1$, then
	\begin{align*}
		\frac{R_{1, a}(1; q^*_a)}{\lambda} & \sim  K(1) \sum_{|v|=N-1} N \frac{(N-1)!}{v!} \frac{\cP(\frac{v}{N})}{v_1 + 1} x^v_N = \frac{K(1)}{x_{1, N}} \sum_{ \substack{|v|=N, \\ v_1 \neq 0} } \frac{N!}{v!} x^v_N \cP(\frac{v}{N}) \\
		& = \frac{K(1)}{x_{1, N}} \Big[ \sum_{ |v|=N } \frac{N!}{v!} x^v_N \cP(\frac{v}{N}) - \sum_{ \substack{|v|=N, \\ v_1 = 0} } \frac{N!}{v!} x^v_N \cP(\frac{v}{N}) \Big].
	\end{align*}
	Here we used the notations $v = (v_1, ..., v_M)$, $|v| = \sum^M_{i=1} v_i$, $v! = v_1! \cdot ... \cdot v_M!$, $x^v_N = x^{v_1}_{1, N} \cdot ... \cdot x^{v_M}_{M, N}$, and $v/N = (v_1/N, ..., v_M/N)$.  
	
	The second quantity in the bracket only includes the situation that no agent selects the lowest spread and thus converges to 0. The first quantity converges to $\cP(x^*)$, by the property of multivariate Bernstein polynomials and the continuity of $\cP(\cdot)$; see a recent survey in \cite{foupouagnigni2020multivariate}.
\end{proof}

\section{Payoff matrix with multiple bid/ask actions and inventory risk}
\begin{figure}[!h]
	\centering
	\includegraphics[width=0.85\linewidth]{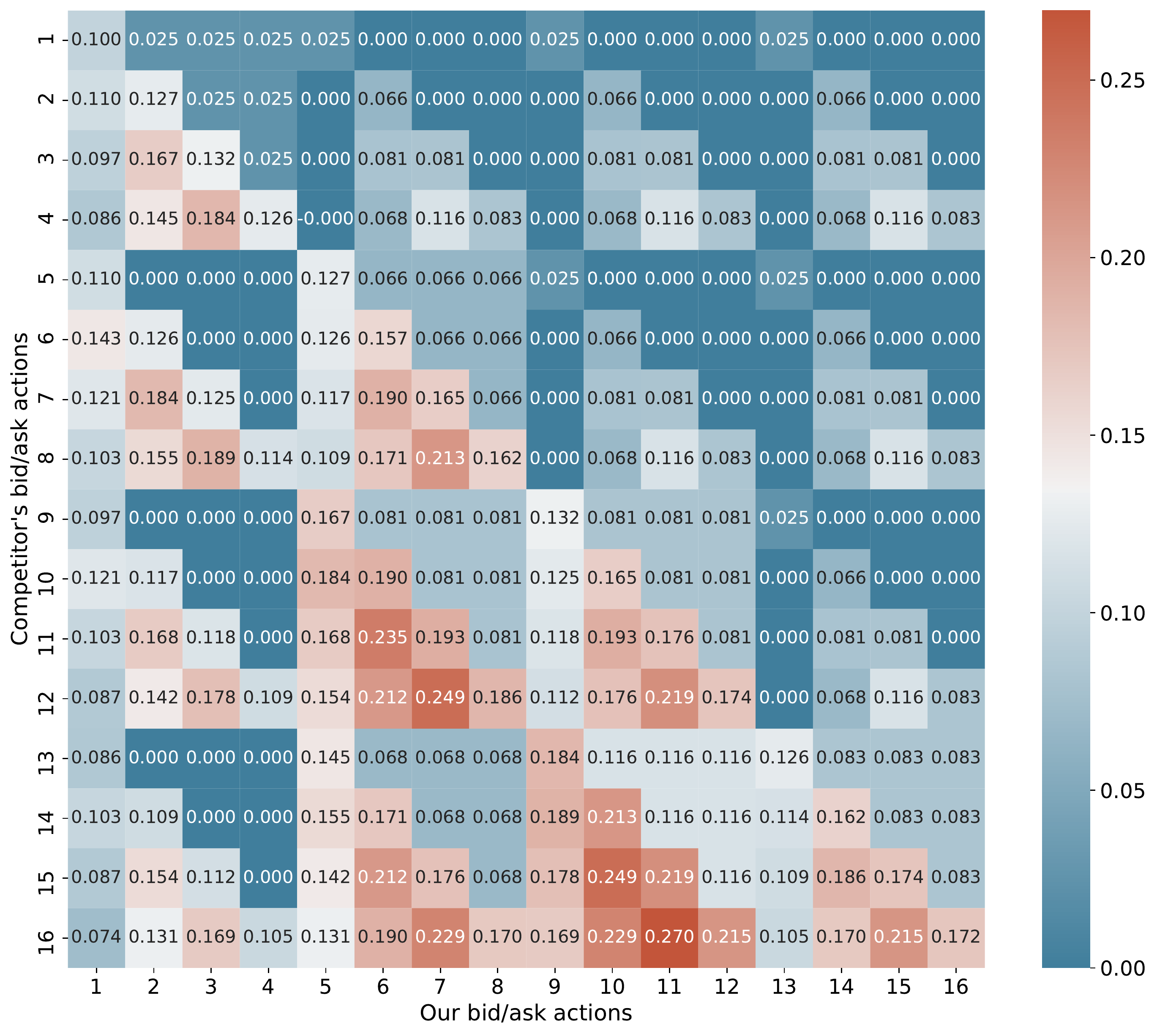}
	\caption{Payoff matrix with multiple bid/ask actions and inventory risk.}
	\label{fig:twosides}
\end{figure}

Figure \ref{fig:twosides} shows the payoff matrix of the second example in Section \ref{sec:multi} for multiple actions. Spreads in each side are selected from $\{3/30, 7/30, 11/30, 15/30\}$. Weights $\{0, 1/30,  2/30, 3/30\}$. Mid-price volatility $\sigma = 0.1$. Inventory risk aversion $\xi = 0.1$. 
\end{document}